\documentclass[traditabstract]{aa} 
\newcommand{\bq}{\begin{eqnarray}}
\newcommand{\eq}{\end{eqnarray}}                                   
\newcommand{\feh}{\element[][]{FeH}}
\newcommand{\teff}{T_{\mathrm{eff}}}

\usepackage{graphicx,natbib,url}
\usepackage{txfonts}
\begin{document}
\title{CRIRES spectroscopy and empirical line-by-line identification of \emph{FeH}
  molecular absorption in an M dwarf\thanks{Data
    were taken at ESO Telescopes under the program 79.D-0357(A)}}

\subtitle{}

\author{S. Wende
       \inst{1}
       \and
       A. Reiners
	  \inst{1}
	  \and
	  A. Seifahrt
	  \inst{2}
	  \and
	  P. F. Bernath
	  \inst{3}
       }

       \institute{Institut f\"ur Astrophysik, Georg-August-Universit\"at
         G\"ottingen, Friedrich-Hund Platz 1, D-37077, Germany\\
         \email{sewende@astro.physik.uni-goettingen.de}
         \email{Ansgar.Reiners@phys.uni-goettingen.de}
         \and
         Physics Department Univ. of California, One Shields
         Avenue Davis, CA 95616 USA \\
         \email{seifahrt@physics.ucdavis.edu}
         \and
          Department of Chemistry, University of York, Heslington, York, YO10 5DD, UK\\
        \email {pfb500@york.ac.uk}}

\date{Received 17 June 2010 / Accepted 22 July 2010}

\abstract{
Molecular FeH provides a large number of sharp and isolated absorption lines that can be used
to measure radial velocity, rotation, or magnetic field strength with high
accuracy.
Our aim is to provide an \feh\ atlas for M-type stars in the spectral
region from $986$\,nm to $1077$\,nm (Wing-Ford band). To identify these lines in
CRIRES spectra of the magnetically inactive, slowly rotating, M5.5 dwarf GJ1002,
we calculated
model spectra for the selected spectral region with theoretical \feh\ line data.
In general this line list agrees with the observed data, but
several individual lines differ significantly in position or in line
strength. After identification of as many as possible
\feh\ lines, we correct the line data for
position and line strength to provide an accurate atlas of \feh\ absorption
lines for use in high precision spectroscopy of low mass stars.
For all lines, we use a Voigt function to obtain their positions and equivalent
widths. Identification with theoretical lines is done by hand.
For confirmation of the identified lines, we use statistical methods,
cross-correlation techniques, and line intensities.
Eventually, we were able to identify \feh\ lines from the $(0,0)$, $(1,0)$, $(1,1)$,
$(2,1)$, $(2,2)$, $(3,2)$, and $(4,3)$ vibrational bands in the
observed spectra and correct the positions of the lines if necessary.
The deviations between theoretical and observed
positions follow a normal distribution approximately around zero. In
order to empirically correct the
line strength, we determined $\teff$, instrumental broadening
(rotational broadening) and a van der Waals enhancement factor for
\feh\ lines in GJ1002. We also give scaling factors for the Einstein
A values to correct the line strengths.
With the identified lines, we derived rotational temperatures from line intensities for GJ1002.
We conclude that \feh\ lines can be used for a wide variety of applications in astrophysics.
With the identified lines it will be possible for example to
characterize magnetically sensitive or very temperature
sensitive lines, which can be used to investigate M-type stars.
}

\keywords{Molecular data - line: identification, profiles - stars: low-mass}

\maketitle
%

\section{Introduction}
High resolution spectroscopy of atomic or molecular lines is used to
measure rotation, magnetic fields, and radial velocity
variations. In solar-like stars, atomic lines are useful to
measure these quantities since they are numerous, well
isolated and sufficiently narrow. In cooler stars, like M
dwarfs, atomic lines are no longer useful because most of them
become very weak, others become
strongly pressure broadened and they are usually overlapped by strong
molecular bands. In these stars molecular lines are a valuable tool to
measure the quantities mentioned above. However, molecular lines tend
to cluster in dense bands e.g. for \element{TiO} and
\element{VO} in the visual
spectral range. Only a few molecules provide absorption lines that are isolated
and can be used for detailed spectroscopic analysis.

The \feh\ molecule provides a particularly large number
of strong and well isolated lines in the z-band ($\sim 990$--$1080$\,nm).
It is the main opacity contributor in this region for late-type dwarf stars,
and can be used for high precision spectroscopy. \feh\ provides
numerous unblended lines that are sufficiently narrow to measure
small broadening effects or variations in the line position.

\citet{1969PASP...81..527W} first discovered the molecular band around
$991$\,nm in the spectra of the cool dwarf Wolf 359. This band was
also found in S-type stars \citep{1972saim.conf..123W} and was
identified as the $(0,0)$ vibrational band of the \feh\ molecule by
\citet{1977A&A....56....1N}. An extensive analysis was carried out by
\citet{1987ApJS...65..721P}. They identified seven vibrational bands
of the $ ^4 \Delta- ^4 \Delta$ electronic transition of the \feh\
molecule and provided tables with molecular constants and quantum
numbers. An important theoretical work, partly based on the previous
one, was carried out by \citet{2003ApJ...594..651D}. They computed a
line list for the $F ^4 \Delta - X ^4 \Delta$ electronic transition
and provided large tables of molecular data with quantum numbers and
line intensities.

\feh\ absorption bands
were also detected in the $J$- and $H$-band with medium resolution spectra
\citep{2003ApJ...582.1066C}. In the $H$-band the $E^4\Pi-A^4\Pi$ electronic
transition is visible \citep{hargreaves2010}.
That \feh\ can be used to determine effective
temperatures was shown for example by \citet{1997ApJ...484..499S} or
\citet{2009A&A...508.1429W}, and its potential
to measure magnetic field strengths was demonstrated by
\citet{2006ApJ...644..497R,2007ApJ...656.1121R}. Theoretical work on the magnetic sensitivity of
\feh\ was published by \citet{2007A&A...473L...1A,2008A&A...482..387A}.

In this paper we use high resolution spectra of GJ1002 to empirically verify
the line list of \citet{2003ApJ...594..651D} and identify \feh\ on a line-by-line basis
in the region $989.8$ $-$ $1076.6$\,nm.
For this purpose we will use Voigt functions to determine the
empirical positions and equivalent widths of the observed \feh\ lines. We identify
the observed lines with theoretical ones by hand, and confirm this identification
using statistical means, cross-correlation techniques, and the line strength
of the identified \feh\ lines. Furthermore, we correct theoretical Einstein A values
to account for mismatches in line depth. We also derive rotational
temperatures from the identified lines, and investigate under which
circumstances they are close to the effective temperature of the star.

\section{Data}
\subsection{CRIRES spectra of GJ1002}
\begin{figure}[!h]
\includegraphics[width=0.45\textwidth,bb = 50 30 540 380]{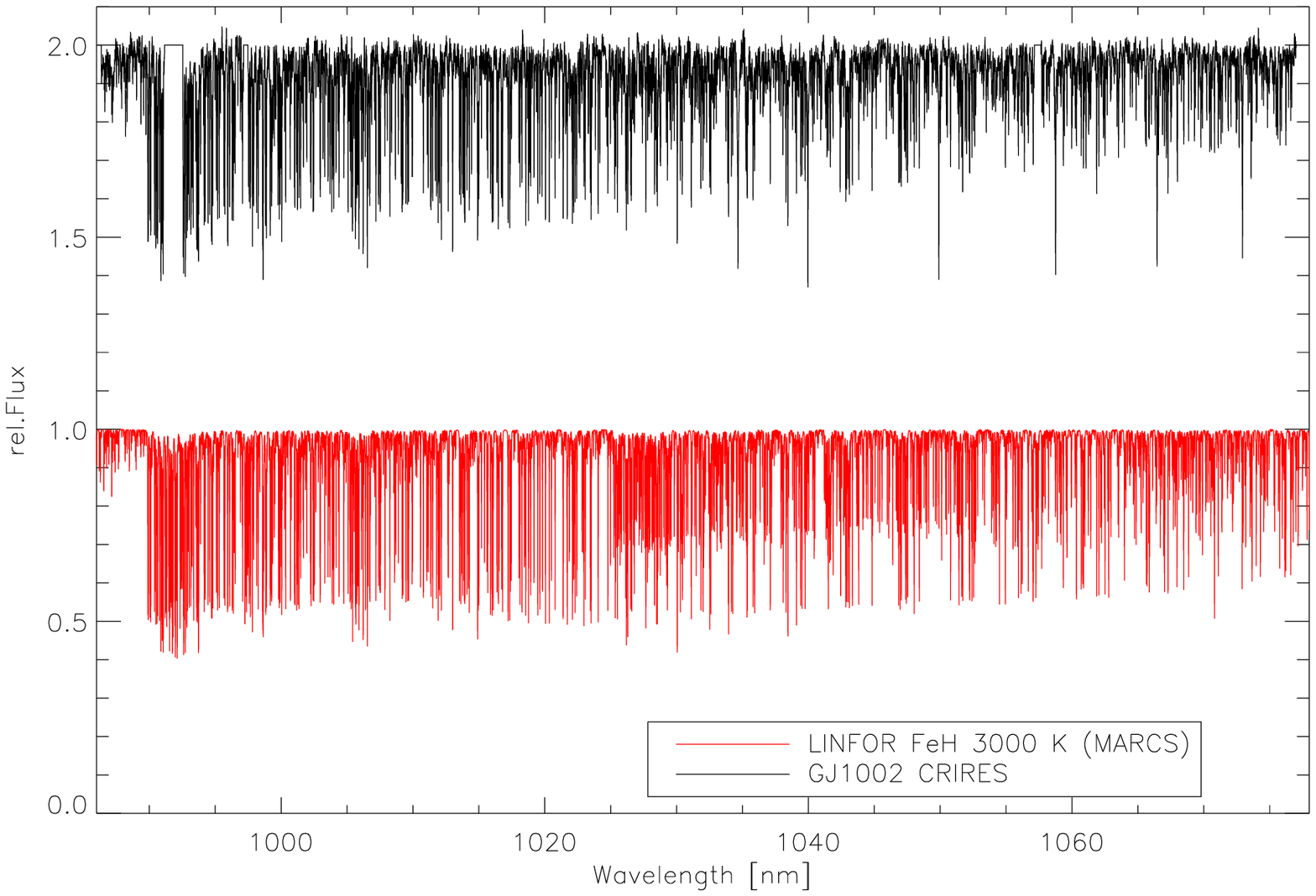}
\includegraphics[width=0.45\textwidth,bb = 50 30 540 380]{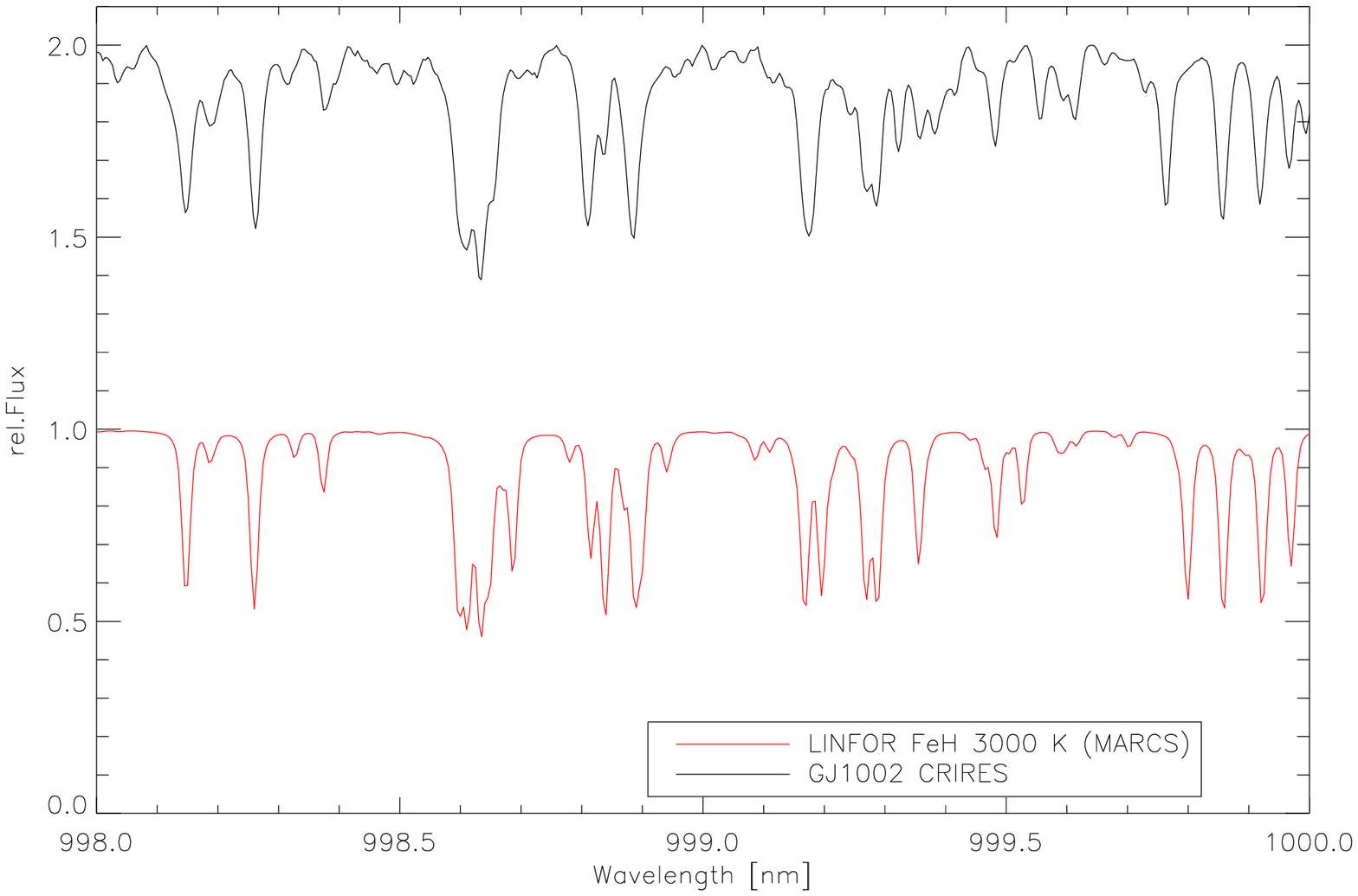}
\caption{top: \feh\ Wing-Ford band, observed (black) and computed (red).
Bottom: a magnification of the spectrum above.}
\label{FeH_obscomp}
\end{figure}

The observational data are CRIRES spectra of the inactive M 5.5 dwarf GJ1002
(see Fig.~\ref{FeH_obscomp}).
The M dwarf has an assumed effective temperature of $3150$\,K (from the spectral
type), and it is a very slow rotator
\citep[$v\sin{i} < 3$\, km s$^{-1}$,][]{2007ApJ...656.1121R}.
There is also very low $H_{\alpha}$ and X-ray activity from
which we can assume that the magnetic field strength is relatively low. Due
to the weak magnetic field and slow rotation, GJ1002 is an ideal
target for identification of molecular \element[][]{FeH} lines. The
lines are only slightly broadened by the different possible mechanisms in contrast to
observations in sun spots where \feh\ lines are always influenced by
a strong magnetic field.

CRIRES observations of GJ1002 were conducted in service mode during
several nights in July 2007. The entrance slit width was set to
0.2\arcsec, hence, the nominal resolving power was $R\sim100\,000$.
Four frames with an integration time of five minutes were
taken in an ABBA nod pattern for each of the nine wavelength settings
covering the region between $986$ -- $1077$\,nm, leaving only one
larger gap at $991.15$ -- $992.45$\,nm and two smaller gaps at
$997.15$ -- $997.50$\,nm and $1057.15$ -- $1057.65$\,nm.

Data reduction made use of the ESOREX pipeline for CRIRES. Science
frames and flatfield frames were corrected for non-linearity and
1D spectra were extracted from the individual flatfielded and sky
subtracted frames using an optimum extraction algorithm.
The wavelength solution is based at first order on the Th-Ar
calibration frames provided by ESO. Due to the slit curvature,
spectra taken in B nodding positions are shifted in wavelength
with respect to spectra taken in A nodding positions. This was
corrected for by mapping all 1D spectra to the spectrum of the
first A nod position.

The individual CRIRES wavelength settings provide a considerable
degree of spectral overlap and up to eight individual spectra were
combined to one final spectrum at each wavelength. While merging
the individual settings, small mismatches in the wavelength solutions
as well as imperfections in the individual spectra (detector cosmetics,
ghost contamination) were corrected. The final spectrum was
normalized to a pseudo-continuum level of unity and finally shifted
to match the McMath FTS spectrum of solar umbra \citep{1998assp.book.....W}.
The error in the wavelength calibration should be smaller than
$0.75$\,km \,s$^{-1}$.

We find the SNR at the continuum level of most parts of the final
spectrum to exceed $200$. The high signal-to-noise ratio and high spectral resolution of the
CRIRES data allow us to identify \element[][]{FeH} lines with high
accuracy.

\subsection{Theoretical \feh\ molecular data and line synthesis}
The theoretical data which we use to identify the \element[][]{FeH} lines is
taken from \citet{2003ApJ...594..651D}. They provide tables of
quantum numbers and energies\footnote{See
\url{http://bernath.uwaterloo.ca/FeH}}. In particular, they provide the vibrational
assignment of the upper and lower states $ v_u$ and $v_l$, respectively,
the projection of the total orbital angular momentum on to the
internuclear axis for the upper and lower state, $ \Omega_u$ and
$ \Omega_l$, and the rotational quantum number$J_l$ for the lower state. Furthermore the
transition branch (P, Q, R), the parity, the wavenumber in cm$^{-1}$, the lower
state energy of the transition $E_l$ and the Einstein A value are given.

From this information it is
possible to estimate $\log_{10}{gf}$-values through \citep{bernath2005}

\bq
 g_lf=\frac{\epsilon_0m_ec^3}{2\pi e^2\nu^2}Ag_u,
\eq
with $ g_l=2J_l+1$ and  $ g_u=2J_u+1$ as the lower and upper statistical weights of
the transition and $\nu=c/\lambda$ as the transition frequency. All quantities
are in SI units.

The van der Waals broadening is determined following
\citet{1996MNRAS.283..821S} which is basically Uns\"old's hydrogenic
approximation. For the ionization energy needed in this approximation
we use an empirically determined value of $6$\,eV which was deduced
from comparison with other diatomic
molecules \citep{2009A&A...508.1429W}. Despite the fact that this is
not the  theoretical value, which is slightly higher, we use this one since
its influence on the van der Waals broadening is not significant.

\begin{figure}[!h]
\includegraphics[width=0.5\textwidth,bb = 15 30 580 800]{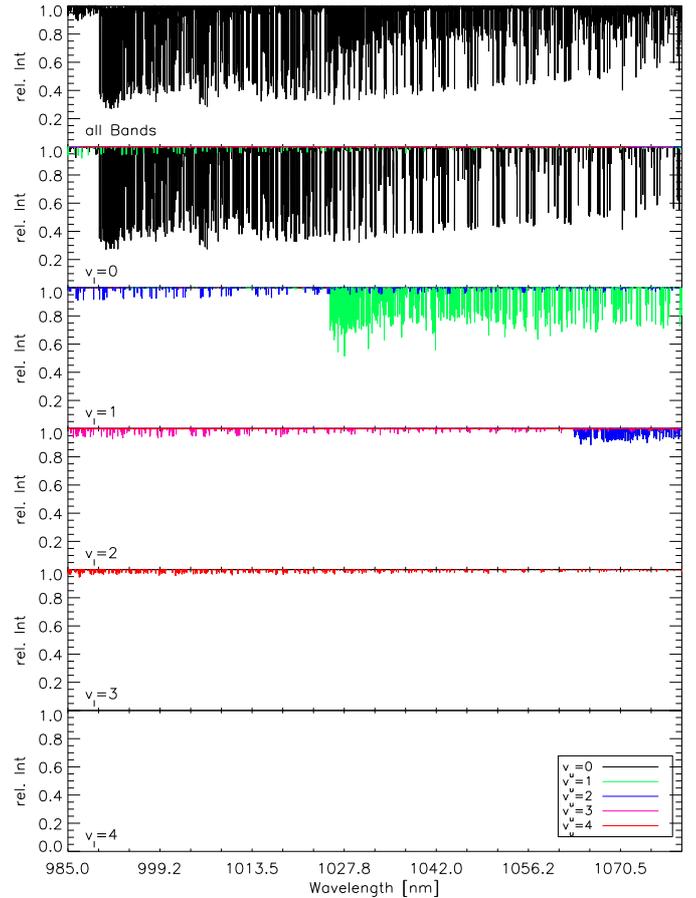}
\caption{\feh\ vibrational bands separated by their vibrational quantum numbers. See color version for more details.}
\label{FeH_all}
\end{figure}

The molecular partition function for \feh\  $Q_{FeH}$, which
is needed for the concentration of \feh\ is computed after
\citet{1984ApJS...56..193S} with molecular data taken from Tables~9 and 10 of
\citet{2003ApJ...594..651D}.
We give in equation~\ref{polypartf} a polynomial expression of a fit to
the partition function which is valid between $1000$\,K and $8000$\,K.
\bq
Q_{FeH}=\sum^4_{i=0}{a_iT^i},
\label{polypartf}
\eq
where $T$ is the temperature in K and
\bq
\left(\begin{array}{r}a_0\\a_1\\a_2\\a_3\\a_4  \end{array}\right)=
\left(\begin{array}{r}-4.9795007e+02\\6.5460944e-01\\3.4171590e-04\\2.7602574e-07\\1.0462656e-11 \end{array}\right),
\label{polycoeff}
\eq
are the coefficients of the polynomial.
For the creation of \feh\  i.e. the concentration, which is governed by
the Saha-Boltzmann equation, we assume $\element{Fe}+\element{H}\rightarrow\feh$.
With this partition function and the data from  \citet{2003ApJ...594..651D}, we
can use a simple description for the absorbance (described in section~\ref{lsc} in
this paper in more detail) to
compute \feh\ spectra with a simple reversing layer model
and separate the lines into vibrational bands in the observed wavelength region
(see Fig.~\ref{FeH_all}). From this figure we can expect that
there will be two sequences of
vibrational transitions present. The sequence with $\Delta v=0$, which
are the $(0,0)$, $(1,1)$, and $(2,2)$ vibrational transitions, and the
sequence with $\Delta v=1$, which are the $(1,0)$, $(2,1)$, $(3,2)$, and
$(4,3)$ transitions. We note that this method does not take into account (among
many other things) the atmospheric structure  or the
chemical composition of the star. It gives only a rough estimate of the relative
strengths of the \feh\ bands.

The full synthetic line formation for the comparison and identification of the observed
\feh\ lines is done with the line
formation code \texttt{SYNTH3} \citep{2007pms..conf..109K}.
This code is able to compute large spectral regions with all \feh\ lines taken from our
line list (see Fig.~\ref{FeH_obscomp}). We use only lines with
$\log_{10}{gf} > -7$
because we assume that other lines have no significant influence. By computing all
the lines in a certain region at once, we account for blends, but we also compute
all lines of the region individually in order to measure their
equivalent width $W_{\lambda}$ using the \texttt{LINFOR3D} code
\citep[based on][]{Bascheck1966}. For the input
model atmospheres we use \texttt{MARCS} \citep{2008A&A...486..951G} with solar
composition \citep{2007SSRv..130..105G}. These models are well suited
for these cool temperatures in low mass stars, since they make use of
up-to-date atomic and molecular data and go down to effective
temperatures of $2500$\,K. We use the plane-parallel, LTE models where the
convection is treated in the mixing-length approximation.  In the
 computation of these model atmospheres, the microturbulence
 parameter was set to zero. However, in the computations of the
 actual spectra, we assume microturbulence parameters according to
 the results of \citet{2009A&A...508.1429W}. We do not expect any
 significant influence since they are on the order of a few hundred
 m s$^{-1}$. We neglect the broadening from macro-turbulent motion,
 which would be hardly visible in the observed spectra.

\section{Methods}
\label{methods}

We start with the investigation
of the observed spectrum, for which we determine the
position of the spectral lines and decide whether a line feature is a
blend or not. Then we measure the equivalent width $W_{\lambda}$ of the lines
with a Voigt fit procedure described below.

We compare the line positions found in the CRIRES data to theoretical
ones and identify them with \feh\ lines. In order to confirm an identification,
we use statistical means:
(i) the method of coincidence, and a cross-correlation technique producing
coincidence curves;
and (ii) a method which takes the intensity into account. In this latter method we will
compare theoretical line strength $S$ (H\"onl-London factor) with observed
$W_{\lambda}$ following \citet{1964BAN....17..311S}.
We also use a description for the absorbance of spectral lines in
order to correct theoretical line intensities given in terms of the Einstein A values.
For this we will compare observed and computed spectra with each other and
obtain a scaling factor for the Einstein A values. The final result is a corrected line list that
reproduces the observed stellar spectrum   as well as the line
positions in sunspot spectra from \citet{1999asus.book.....W}. The
line intensities are hard to confirm in the solar case, since many \feh\ lines
are strongly split by magnetic fields.

\subsection{Voigt Fit}
\label{mvf}
In order to measure $W_{\lambda}$ in the observed spectra we use a
`multi-Voigt fit' procedure (based on IDL curvefit function). The `multi-Voigt fit'
is defined as
\bq
F=\sum{}_{i=0}^{N}A_i\cdot \frac{H(u_i,a_i)}{max(H(u_i,a_i))},
\eq
where A is the amplitude describing the depth of the line,
\bq
u=\frac{\lambda-\lambda_0}{\sigma},
\label{doppler_u}
\eq
with $\sigma$ as Gaussian (or Doppler) width,
\bq
a=\frac{\gamma}{4\pi}
\frac{{\lambda_0}^2}{c}\frac{1}{\sigma},
\label{voigt_a}
\eq
 which will be called the Voigt constant throughout in this paper, $\gamma$ is the radiation damping constant,
and $H(u,a)$ is the Hjerting function \citep[][and references therein]{2008oasp.book.....G}.
We successively fit the
whole spectrum within bins of $7$\,\AA\ . The first and the last $1$\,\AA\ are
cut off after the fit to avoid boundary effects and we use only line
profiles whose centers are inside the $5$\,\AA\ bin. For the Voigt
profiles inside a bin, we assume a constant $a$-value for the Lorentz part and a fixed Gaussian
width, but we allow changes in these two parameters from one wavelength bin to
another because the \feh\ lines tend to become narrower towards longer
wavelengths. This is probably because we use $a$ and $\sigma$ only as fit parameters, and
\feh\ lines starts to saturate at the band head, but become weaker towards
longer wavelengths. Hence, the width of the line profile, which is a combination of $a$ and
$\sigma$ decreases with decreasing saturation and so both parameters
decrease.
 The wavelength dependence of the Doppler width affects the
width of the lines as well, but it is negligible and goes in
opposite direction (for example,
$\sigma=0.1$\,\AA\ at $\lambda=10\,000$\,\AA\ would change to
$\sigma=0.105$\,\AA\ at $\lambda=10\,500$\,\AA\ ).

\begin{figure}[!h]
\includegraphics[width=0.45\textwidth,bb = 90 30 540 380]{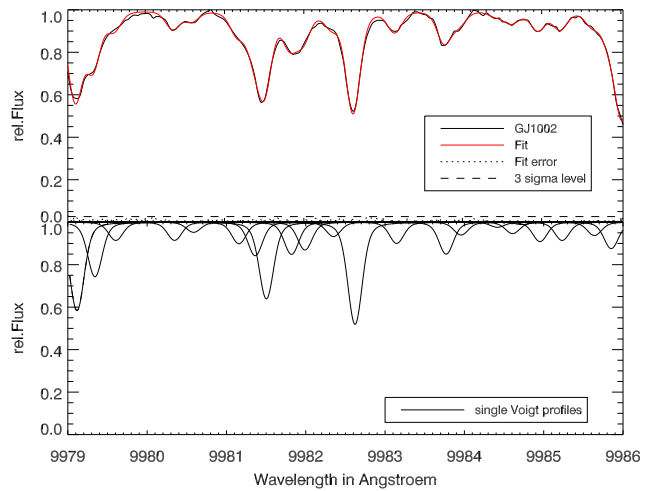}
\caption{Upper panel: Part of the observed spectrum of GJ1002 with multiple Voigt fit.
Lower panel: Single Voigt functions which were returned by the fit.}
\label{voigtfit}
\end{figure}

With this fit procedure, we obtain the parameters (position,
depth (amplitude), Voigt constant, and $\sigma$), needed for the individual Voigt line
profiles to fit the observed spectrum. In Fig.~\ref{voigtfit}, we show
an example of how the fit (red) reduced the observed spectrum (black) into single
line profiles (bottom panel).

The convergence criterion is the minimization of the residual
flux between observation and fit ($O-C$). The iteration is stopped if the
maximal error of the fit is lower than three times the standard
deviation of the error or if the standard deviation of the error does
not change significantly between two iterations.
From these Voigt profiles, $W_{\lambda}$ can easily be computed by integrating
over the single line profiles.
The fit is also able to find and separate possible blends. For this, we assume
that the blended components differ in position by at least $0.1$\,\AA\ .

The measured $W_{\lambda}$ will be assigned to the associated
theoretical lines. This means if an observed line can be identified
with exactly one theoretical line, $W_{\lambda}$ is fully assigned to this one theoretical
line. However, in most cases theory predicts more than one line for an
observed line position because of numerous overlapping vibrational bands
in the same wavelength region and from many closely-spaced line pairs which only differ in parity.
 In these cases, $W_{\lambda}$ will be distributed to
all predicted theoretical lines at this position. This is done
using the ratio
$R_i$ of theoretical equivalent widths $^{theo}W_{\lambda}^i$ obtained
from the individual theoretical line profiles:
\bq
R_i=\frac{^{theo}W_{\lambda}^i}{\sum_{i \in B}~ ^{theo}W_{\lambda}^i},
\label{EW_ratio}
\eq
with $B$ as the set of blended components.
There is of course the chance that \feh\ is not the only
contributor to the observed blend. To avoid this uncertainty, we
additionally use the line intensities in a method described below.

\subsection{Method of coincidence}
In order to confirm the identification of a molecular band, or a
sub-band, we use the `method of coincidence', which was first introduced
by \citet{1929ApJ....69..196R}. It gives the number of lines $C$ in a spectral
range that will be found by chance if one uses a set of randomly generated line
positions and compare them with observed ones. Fundamental probability calculations
lead to
\bq
C=M[ 1-\exp{(-2xw)}],
\label{moc}
\eq
where $M$ is the number of lines in a particular region, $x$ is the tolerated deviation
in position, and $w$ is the line density
(average number of spectral lines in the investigated region). 
 This means that $C/M=p_{\rm random}$ gives the probability of finding,
with a randomly chosen
line position, a random coincidence.
If one identifies $N$ out of $M$ lines in the observations, then the probability
of finding an identification is $N/M=p_{\rm identified}$. The ratio $N/C$
describes how likely it is to find per identified line a line by
random coincidence.
Hence, the number of actually identified lines $N$ should exceed the number of
$C$ of purely random coincidences. If this is the case, then one can
assume that the lines are probably identified in the observed spectra.
\subsection{Theoretical Line Strength}
\label{tls}
For the identification of molecular lines, it is also useful to take
the intensities of the lines into account.
In order to compare the theoretical line strength $S$ with the observed
equivalent width $W_{\lambda}$, we
follow \citet{1964BAN....17..311S}.
For weak lines (mildly saturated) $W_{\lambda}$ is proportional to the
wavelength $\lambda_0$, the oscillator strength $f$, and $N_i$, the number of
absorbers \citep{2008oasp.book.....G}:
\bq
W_{\lambda} \propto \lambda_0^2fN_i,
\label{W}
\eq
and
\bq
 f \propto \frac{g_u}{g_l}\lambda_0^2A,
\label{f}
\eq
with $N_i$ as the number of absorbers given by
the Boltzmann distribution
\bq
 N_i \propto g_le^{(\frac{-hc}{kT}E_0)},
\label{N_i}
\eq
where $E_0$ is the lower state energy  level in cm$^{-1}$.
In order to derive the final form, we use equation~\ref{f} and
\ref{N_i} together with equation~\ref{W}
and the expression for the Einstein A value
\bq
 A=\frac{A_{v_u,v_l}S}{g_u}.
\eq
Here $ A_{v_u,v_l}$ is the Einstein $A$ for a specific vibrational transition
and constant for the vibrational band. The resulting equation is
\bq
\log_{10}{\frac{W_{\lambda}}{S\lambda_0^4}}=C-\frac{hc}{kT}E_0\log_{10}{e},
\label{logS}
\eq
which is the relation from \citet{1964BAN....17..311S} with $E_0$
replacing  $ BJ_l(J_l+1)$. 
 Possible stimulated emission can be neglected for our analysis since
there is only a small population in the excited state.
If we plot $\log_{10}{W_{\lambda}/S\lambda_0^4}$ against
$E_o$, a straight line should be found, which then
suggests the correct identification
of the lines in the \feh\ molecular band.
\begin{figure}[!h]
\includegraphics[width=0.45\textwidth,bb = 30 30 540 380]{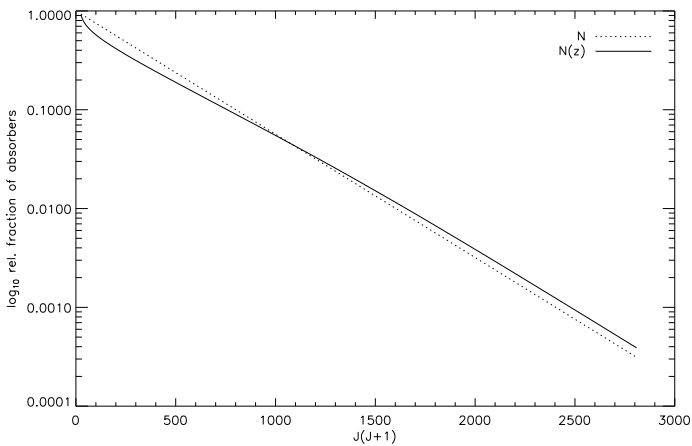}
\caption{Relative number of absorbers with (dashed line), and without
 (solid line) constant number of \feh\ molecules over different
 atmospheric layers.}
\label{varN}
\end{figure}
Equation~\ref{logS} accounts for different branches as well as for
different $\Omega$. This is because $S$ on the  left side of
equation~\ref{logS} is computed for all branches (P, Q, R) and $\Omega$ values.

The $S$ values that we will use in the analysis are computed for the
intermediate Hund's case and are determined from the Einstein $A$ values given by
\citet{2003ApJ...594..651D}.

We point out that we implicitly assumed in equation~\ref{N_i}
that the total number of \feh\ molecules $N$ is
constant for all lines. This assumption is only valid if we consider a
small  isothermal atmospheric layer. However, since a spectral
line forms over several layers,
this assumption is not exactly valid. If we furthermore consider a set
of lines, with a wide range in $gf$ and $E_0$ values, then we expect
that these lines form at different heights. Thus, the number of \feh\
molecules is not constant anymore: strong spectral lines are assumed
to be formed in higher atmospheric layers, and weak lines in deeper
layers.  Hence, deviations from a straight
line in the $[\log_{10}{W_{\lambda}}/S,E_0]$ diagram for lines with
very small and very high $J$ are expected.  Assuming more \feh\ molecules
in deeper layers, due to higher density,
larger equivalent widths for these lines can be expected.
For a qualitative description, we assume the total number of \feh\
molecules to be inversely proportional to the equivalent width, which
reflects  the heights of formation for weak and strong lines.
\bq
N(z)=N\cdot \alpha W_{\lambda}^{\beta}.
\label{Nvar}
\eq
$\alpha$ and $\beta$ are free parameters and here chosen as $1.25$ and
$-0.25$, respectively. We plot the right hand side of equation~\ref{logS} for
constant and variable molecule number in Fig.~\ref{varN}. However, the
situation is much more complicated and we use synthetic line formation
to investigate this behavior in the results section.

Following \citet{1964BAN....17..311S}, we can now use the
$[\log_{10}{W_{\lambda}}/S,E_0]$ diagram to classify the
identified \feh\ lines into one of the following classes:
\begin{enumerate}
\item P - the line is present, and its $W_{\lambda}$ agrees well with the straight
line of the diagram.

\item Pb - the line is present, but its $W_{\lambda}$ is too large, i.e. the
$\log_{10}{W_{\lambda}}/S$ value lies above the straight line of the diagram. This
could imply that the line is blended by an other element (or that its
computed line strength is too small).

\item R - the line strength is presumably reduced by
 perturbations. That means, that the computed line strength is too
 large, and the data point lies below the line.

\item Q - we identified the line, but we can't verify its
 identification, because we only investigate lines with $\Delta
 \Omega=0$ in this plot.\\
\end{enumerate}
Eventually, it should be possible to derive the excitation
temperature for the rotational transitions $T$ in equation \ref{logS} from the slope
of a linear fit in the $[\log_{10}{W_{\lambda}}/S,E_0]$
diagram. However,  one has to be careful.
\citet{1970SoPh...15..342W} reports,  and we experienced difficulties, that the obtained rotational temperature from
this method relies crucially on the data points which are
included in the linear fit and also on the degree of accuracy in
measuring $W_{\lambda}$.

\subsection{Line strength correction}
\label{lsc}
In some cases $W_{\lambda}$ and line depths of the observed \feh\ lines do not match the
computed ones from the line list. In general, we observe that the
differences increase towards longer wavelengths  and computed lines
become stronger than observed ones.
 The strength of the lines are mainly determined by the lower
state energies $E_0$ and Einstein A values. We will correct only the Einstein A values, since a set of
high resolution spectra in the z-band for different temperatures would
be required to correct $E_0$.
In order to correct the Einstein A value, we use the formula from
\citet{bernath2005} for the absorbance
in a modified form in which the Einstein A values enter the expression
\bq
 -\ln\left({\frac{I}{I_0}}\right)_x=\frac{(2J_u+1)A_x}{8\pi\bar{\nu}^2q}e^{-E_0^x/kT}(1-e^{-h\nu/kT})GNl,
\label{absorbance}
\eq
where $\bar{\nu}$ is the frequency of a molecular in cm\,$^{-1}$, N is the number of molecular
absorbers per cubic cm in the energy state (population density), $q$ is the partition function, l is the
length, e.g. for an atmospheric
layer, and $G$ is a line profile function, e.g. a Voigt  function.
If we compare the observed spectra with the computed one (produced
with the \texttt{SYNTH3} code) and assume that both atmospheres
have the same structure, then $A_x$ and $E_0^x$ are the only variables that account for differences in the
spectra ($x$ stands for either the computed or observed spectra).
 We assume also $E_0^x$ as constant and write
\bq
\frac{-\ln\left({\frac{I}{I_0}}\right)_{obs}}{-\ln\left({\frac{I}{I_0}}\right)_{comp}}=\frac{A_{obs}}{A_{comp}}.
\label{intdiff}
\eq
 This shows that the correction for the Einstein A value is just a
linear scaling with the ratio of the intensities
\bq
A_{obs}=\frac{-\ln\left({\frac{I}{I_0}}\right)_{obs}}{-\ln\left({\frac{I}{I_0}}\right)_{comp}}A_{comp}=s A_{comp}.
\label{Ascale}
\eq
We introduced the scaling factor $s$  for the ratio of both
intensities. If we use that  $ A=A_{v_u,v_l}S/(2J_u+1)$
\citep{2003ApJ...594..651D}, then
\bq
 s=\frac{A_{v_u,v_l}^{obs}}{A_{v_u,v_l}^{comp}}\frac{S_{obs}}{S_{comp}},
\label{SA}
\eq
where $S$ is the H\"onl-London factor. If the theoretical $S$ values
are wrong, we could perhaps detect it using
equation~\ref{logS} or perhaps in a possible dependence of the scaling factors
on  $ J_l$.

\section{Results}
\subsection{Atomic line identification and unidentified lines}
Since atomic lines are present in the wavelength region of the
observed spectra of GJ1002, we use the VALD
\footnote{\url{http://vald.astro.univie.ac.at}}
\citep{2000BaltA...9..590K} database to include the available atomic
line data in our calculations. We give a list of atomic
lines which are present at $\teff=3100$\,K in
Table~\ref{Atomic_table}.
We included lines with line depths deeper than
$2\%$ below the continuum in the computed spectra. We do not try to
correct the atomic line positions or line strengths
and take the data as provided by VALD.

After we identify the atomic lines and \feh\ lines, there are still
unidentified lines which seem to belong to neither \feh\ nor to a known atomic line.
We give a list of these lines for which the line depth is deeper
than $10\%$ below the continuum (Table~\ref{unidlines}).
Our opinion is that most of these lines belong to
\feh\, but we were not able to identify them with confidence.
\begin{table*}
\centering
\caption{Identified atomic lines}
\begin{tabular}{cccc|cccc|cccc}

\hline\hline
$Ion$ & $Position [\AA\ ]$ & $E_l$& $\log{gf}$ &   $Ion$ & $Position [\AA\ ]$ & $E_l$& $\log{gf}$ & $Ion$ & $Position [\AA\ ]$ & $E_l$& $\log{gf}$\\
\hline
'Cr1' &        9903.6226 & 2.987 & -2.131 & 'Fe1' &        10116.787 & 2.759 & -3.705 & 'Ti1' &        10399.651 & 0.848 & -1.623\\
'Ti1' &        9930.0728 & 1.879 & -1.580 & 'Ti1' &        10123.668 & 2.175 & -1.722 & 'Cr1' &        10419.476 & 3.013 & -1.806\\
'Ti1' &        9944.1036 & 2.160 & -1.821 & 'Ti1' &        10148.300 & 3.148 & -0.910 & 'Fe1' &        10425.885 & 2.692 & -3.627\\
'Ti1' &        9951.7317 & 2.154 & -1.778 & 'Fe1' &        10148.342 & 4.796 & -0.177 & 'Ti1' &        10462.915 & 2.256 & -2.054\\
'Cr1' &        9951.7997 & 3.556 & -1.129 & 'Fe1' &        10157.947 & 2.176 & -4.225 & 'Fe1' &        10472.522 & 3.884 & -1.187\\
'Ti1' &        9967.4679 & 1.053 & -4.108 & 'Fe1' &        10170.256 & 2.198 & -4.114 & 'Cr1' &        10489.119 & 3.011 & -0.972\\
'Ti1' &        10000.700 & 1.873 & -1.840 & 'Ti1' &        10173.274 & 1.443 & -3.465 & 'Ti1' &        10498.990 & 0.836 & -1.739\\
'Ti1' &        10005.828 & 2.160 & -1.124 & 'Fe1' &        10197.900 & 2.728 & -3.589 & 'Cr1' &        10512.887 & 3.013 & -1.558\\
'Ti1' &        10008.403 & 1.067 & -3.626 & 'Ca1' &        10201.981 & 4.680 & -0.369 & 'Fe1' &        10535.121 & 3.929 & -1.482\\
'Ti1' &        10014.490 & 2.154 & -1.284 & 'Ca1' &        10205.743 & 4.681 & -0.199 & 'Cr1' &        10552.983 & 3.011 & -1.976\\
'Sc1' &        10027.787 & 1.865 & -1.286 & 'Ca1' &        10205.803 & 4.681 & -1.102 & 'Ti1' &        10554.648 & 1.887 & -2.607\\
'Ti1' &        10037.242 & 1.460 & -2.227 & 'Ca1' &        10211.457 & 4.681 & -0.039 & 'Fe1' &        10580.037 & 3.301 & -3.137\\
'Ti1' &        10051.581 & 1.443 & -2.205 & 'Ca1' &        10211.561 & 4.681 & -1.102 & 'Ti1' &        10587.533 & 0.826 & -1.866\\
'Sc1' &        10060.261 & 1.851 & -1.479 & 'Fe1' &        10221.204 & 3.071 & -2.760 & 'Ti1' &        10610.624 & 0.848 & -2.761\\
'Fe1' &        10060.397 & 5.033 & -1.231 & 'Fe1' &        10268.031 & 2.223 & -4.533 & 'Fe1' &        10619.630 & 3.267 & -3.128\\
'Ti1' &        10060.485 & 2.175 & -0.894 & 'Ca1' &        10291.397 & 4.624 & -0.265 & 'Cr1' &        10650.558 & 3.011 & -1.613\\
'Ti1' &        10062.662 & 1.430 & -2.351 & 'Fe1' &        10343.720 & 2.198 & -3.574 & 'Ti1' &        10664.544 & 0.818 & -2.007\\
'Fe1' &        10067.804 & 4.835 & -0.288 & 'Ca1' &        10346.655 & 2.933 & -0.408 & 'Cr1' &        10670.437 & 3.013 & -1.489\\
'Ti1' &        10069.273 & 2.160 & -1.750 & 'Fe1' &        10350.802 & 5.393 & -0.548 & 'Cr1' &        10675.063 & 3.013 & -1.374\\
'Ti1' &        10077.885 & 1.067 & -4.065 & 'Fe1' &        10381.844 & 2.223 & -4.145 & 'Ti1' &        10679.972 & 0.836 & -2.592\\
'Cr1' &        10083.115 & 3.556 & -1.307 & 'Ti1' &        10393.591 & 1.502 & -2.595 & 'Fe1' &        10728.124 & 3.640 & -2.763\\
'Fe1' &        10084.158 & 2.424 & -4.544 & 'Cr1' &        10394.793 & 3.010 & -2.006 & 'Ti1' &        10729.329 & 0.813 & -2.156\\
'Cr1' &        10114.770 & 3.013 & -2.073 & 'Fe1' &        10398.645 & 2.176 & -3.390 & 'Ca1' &        10729.654 & 4.430 & -1.841\\

\hline
\end{tabular}
\label{Atomic_table}
\end{table*}

\begin{table*}
\centering
\caption{List of unidentified lines deeper then $0.9$.}
\begin{tabular}{cc|cc|cc|cc|cc|cc|cc}
\hline\hline
$\lambda_{vac}$ [\AA\ ] & Depth & $\lambda_{vac}$ [\AA\ ] & Depth  & $\lambda_{vac}$ [\AA\ ] & Depth & $\lambda_{vac}$ [\AA\ ] & Depth & $\lambda_{vac}$ [\AA\ ] & Depth & $\lambda_{vac}$ [\AA\ ] & Depth & $\lambda_{vac}$ [\AA\ ] & Depth   \\
\hline
     9904.46 & 0.88 &       10084.8 & 0.90 &       10235.9 & 0.84 &       10344.9 & 0.86 &       10424.9 & 0.79 &       10547.5 & 0.80 &       10644.2 & 0.84\\
     9909.42 & 0.87 &       10095.7 & 0.86 &       10238.5 & 0.73 &       10346.0 & 0.76 &       10427.8 & 0.89 &       10548.2 & 0.85 &       10645.7 & 0.84\\
     9927.74 & 0.79 &       10098.8 & 0.89 &       10242.9 & 0.89 &       10347.3 & 0.82 &       10438.8 & 0.87 &       10548.7 & 0.89 &       10655.3 & 0.80\\
     9930.70 & 0.81 &       10107.3 & 0.85 &       10246.4 & 0.81 &       10347.5 & 0.87 &       10440.0 & 0.85 &       10550.8 & 0.88 &       10660.2 & 0.79\\
     9931.26 & 0.87 &       10107.6 & 0.86 &       10247.3 & 0.74 &       10347.7 & 0.87 &       10442.0 & 0.80 &       10555.8 & 0.90 &       10660.7 & 0.87\\
     9932.82 & 0.88 &       10108.7 & 0.89 &       10247.3 & 0.74 &       10351.3 & 0.89 &       10443.2 & 0.84 &       10559.0 & 0.89 &       10665.1 & 0.84\\
     9932.94 & 0.88 &       10116.9 & 0.65 &       10248.4 & 0.84 &       10354.4 & 0.79 &       10443.6 & 0.89 &       10563.5 & 0.82 &       10665.6 & 0.71\\
     9938.22 & 0.88 &       10140.2 & 0.89 &       10255.4 & 0.84 &       10357.5 & 0.89 &       10444.6 & 0.90 &       10563.8 & 0.67 &       10666.9 & 0.87\\
     9978.82 & 0.89 &       10141.5 & 0.73 &       10260.6 & 0.88 &       10358.5 & 0.74 &       10444.8 & 0.81 &       10565.8 & 0.82 &       10669.6 & 0.88\\
     9983.94 & 0.90 &       10144.5 & 0.66 &       10261.0 & 0.88 &       10361.5 & 0.86 &       10444.9 & 0.90 &       10567.1 & 0.83 &       10671.3 & 0.90\\
     9984.94 & 0.90 &       10164.4 & 0.89 &       10266.4 & 0.81 &       10363.3 & 0.89 &       10446.7 & 0.88 &       10568.8 & 0.90 &       10671.6 & 0.71\\
     9985.22 & 0.90 &       10167.5 & 0.84 &       10267.3 & 0.79 &       10367.9 & 0.87 &       10449.3 & 0.88 &       10571.1 & 0.89 &       10673.3 & 0.88\\
     9993.22 & 0.72 &       10171.2 & 0.89 &       10274.1 & 0.74 &       10369.1 & 0.88 &       10463.2 & 0.81 &       10577.5 & 0.90 &       10674.7 & 0.83\\
     9993.82 & 0.77 &       10174.1 & 0.89 &       10280.7 & 0.87 &       10378.0 & 0.83 &       10463.9 & 0.84 &       10580.6 & 0.85 &       10675.5 & 0.81\\
     9994.14 & 0.87 &       10177.2 & 0.90 &       10281.0 & 0.77 &       10381.0 & 0.88 &       10464.3 & 0.77 &       10582.8 & 0.80 &       10681.4 & 0.90\\
     9997.30 & 0.88 &       10187.0 & 0.81 &       10286.1 & 0.89 &       10382.0 & 0.90 &       10464.9 & 0.90 &       10586.5 & 0.87 &       10697.2 & 0.89\\
     9999.94 & 0.77 &       10188.6 & 0.67 &       10287.4 & 0.82 &       10382.7 & 0.78 &       10477.6 & 0.75 &       10586.9 & 0.75 &       10697.3 & 0.88\\
     10005.3 & 0.89 &       10193.7 & 0.83 &       10297.2 & 0.80 &       10383.8 & 0.85 &       10479.6 & 0.86 &       10588.5 & 0.90 &       10711.9 & 0.81\\
     10015.0 & 0.81 &       10197.2 & 0.87 &       10298.0 & 0.88 &       10383.9 & 0.85 &       10490.9 & 0.83 &       10589.6 & 0.89 &       10712.3 & 0.86\\
     10021.2 & 0.88 &       10200.1 & 0.89 &       10303.5 & 0.88 &       10385.7 & 0.89 &       10493.9 & 0.87 &       10593.8 & 0.84 &       10713.0 & 0.74\\
     10023.7 & 0.87 &       10202.9 & 0.90 &       10309.6 & 0.86 &       10387.8 & 0.86 &       10499.5 & 0.83 &       10594.3 & 0.76 &       10716.6 & 0.85\\
     10027.7 & 0.89 &       10203.0 & 0.90 &       10310.6 & 0.84 &       10392.2 & 0.87 &       10499.6 & 0.83 &       10600.7 & 0.87 &       10724.6 & 0.77\\
     10029.3 & 0.86 &       10209.1 & 0.86 &       10312.2 & 0.75 &       10396.0 & 0.73 &       10500.2 & 0.75 &       10601.8 & 0.89 &       10725.3 & 0.81\\
     10031.8 & 0.70 &       10211.1 & 0.70 &       10312.5 & 0.89 &       10398.4 & 0.89 &       10500.9 & 0.87 &       10602.9 & 0.88 &       10726.4 & 0.86\\
     10035.9 & 0.83 &       10214.5 & 0.79 &       10314.7 & 0.84 &       10399.2 & 0.87 &       10514.7 & 0.84 &       10604.0 & 0.89 &       10734.7 & 0.85\\
     10041.9 & 0.90 &       10215.8 & 0.77 &       10319.3 & 0.90 &       10400.2 & 0.82 &       10515.1 & 0.88 &       10605.3 & 0.90 &       10739.2 & 0.90\\
     10048.2 & 0.82 &       10217.8 & 0.89 &       10319.5 & 0.85 &       10400.5 & 0.88 &       10515.5 & 0.79 &       10605.5 & 0.88 &       10740.5 & 0.82\\
     10051.1 & 0.86 &       10219.5 & 0.86 &       10320.6 & 0.86 &       10402.8 & 0.83 &       10516.6 & 0.89 &       10609.5 & 0.87 &       10744.1 & 0.86\\
     10066.5 & 0.88 &       10219.9 & 0.87 &       10322.2 & 0.88 &       10406.5 & 0.89 &       10519.3 & 0.88 &       10609.6 & 0.88 &       10745.3 & 0.84\\
     10068.7 & 0.79 &       10220.0 & 0.88 &       10322.8 & 0.75 &       10406.8 & 0.81 &       10519.7 & 0.89 &       10612.7 & 0.77 &       10748.7 & 0.90\\
     10073.9 & 0.58 &       10223.9 & 0.85 &       10327.8 & 0.89 &       10408.4 & 0.87 &       10532.9 & 0.86 &       10613.1 & 0.82 &       10749.4 & 0.88\\
     10074.5 & 0.89 &       10225.8 & 0.77 &       10327.9 & 0.87 &       10409.3 & 0.75 &       10535.5 & 0.89 &       10614.0 & 0.81 &       10752.1 & 0.90\\
     10075.9 & 0.90 &       10226.7 & 0.76 &       10328.3 & 0.78 &       10410.1 & 0.81 &       10536.2 & 0.86 &       10626.7 & 0.82 &       10753.7 & 0.90\\
     10078.5 & 0.86 &       10227.7 & 0.88 &       10332.7 & 0.88 &       10415.7 & 0.89 &       10539.5 & 0.83 &       10629.9 & 0.88 &       10754.7 & 0.79\\
     10080.6 & 0.75 &       10229.4 & 0.89 &       10338.9 & 0.64 &       10417.8 & 0.88 &       10543.1 & 0.87 &       10634.8 & 0.89 &       10755.8 & 0.90\\
     10083.8 & 0.72 &       10230.5 & 0.73 &       10340.0 & 0.89 &       10422.5 & 0.90 &       10546.1 & 0.86 &       10638.3 & 0.80 &       10759.4 & 0.89\\
\hline
\end{tabular}
\label{unidlines}
\end{table*}

\subsection{\feh\ line identification}

\begin{figure}[!h]
\includegraphics[width=0.45\textwidth,bb = 30 30 540 380]{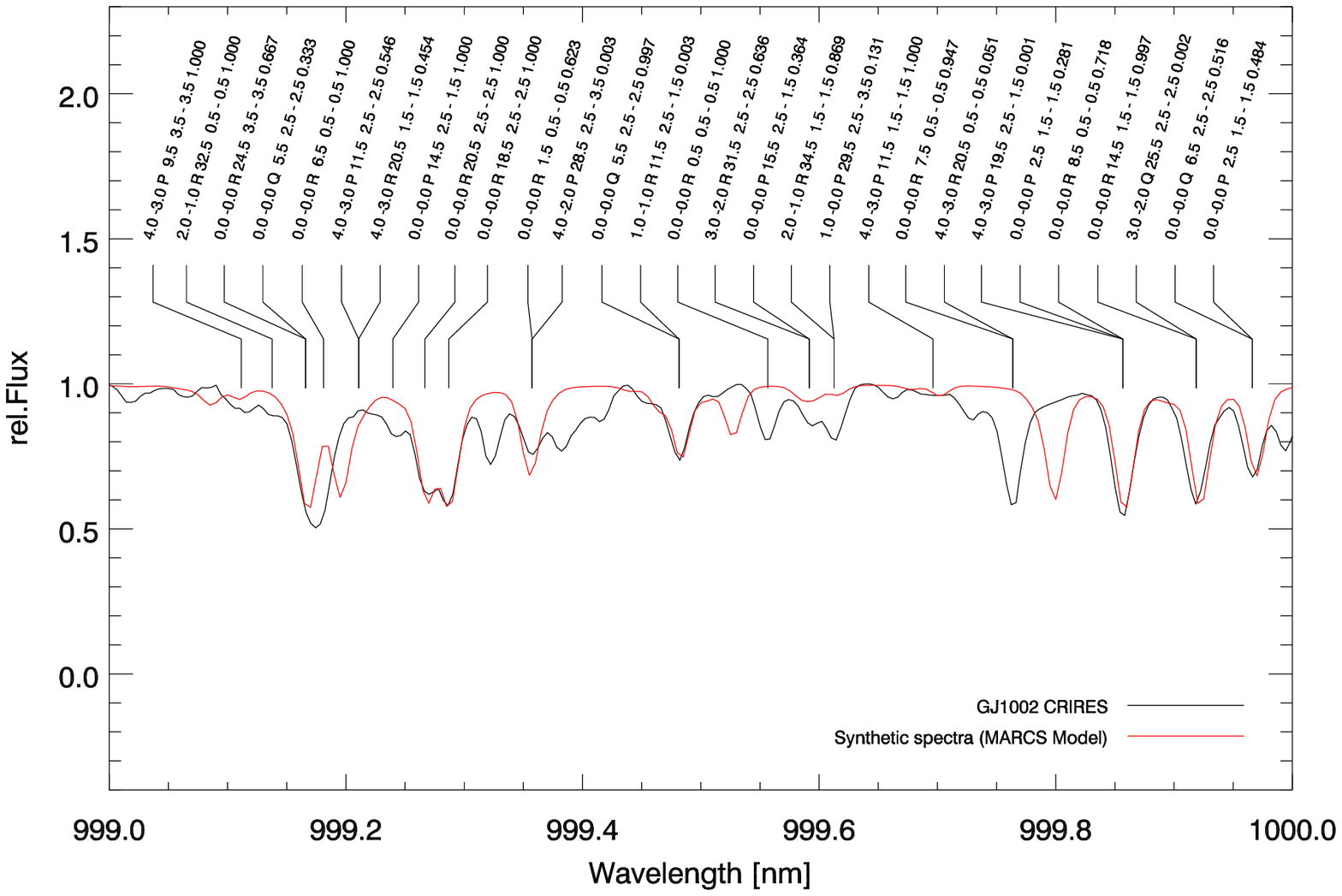}
\caption{Observed spectrum of GJ1002 (black) and computed one (red) labeled with quantum numbers.}
\label{ident}
\end{figure}

\begin{figure}[!h]
\includegraphics[width=0.45\textwidth]{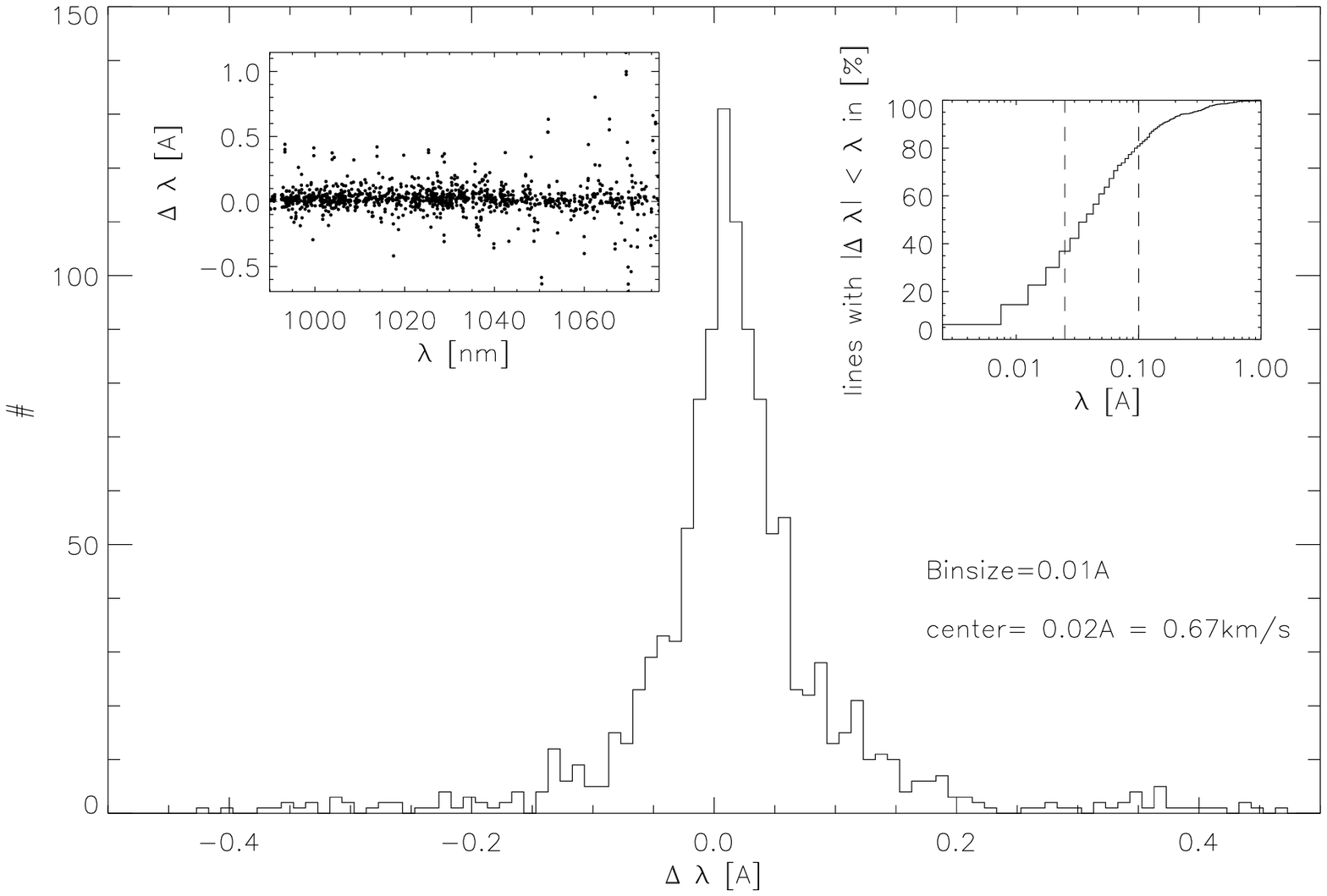}
\caption{ Histogram of the residuals between computed line positions
and observed ones.
In the upper left corner, the residuals are plotted against wavelength.
In the upper right corner, the fraction of lines with residuals lower
than a certain value are plotted. The left dashed line represents the
accuracy of the wavelength solution which is $0.025$\,\AA\, the right one the detection
boundary of $0.1$\,\AA\ .}
\label{deltalam}
\end{figure}

We use the results from the Voigt fit to assign the individual Voigt profiles
from the observed lines to the individual theoretical \feh\ lines. We did this by
hand, and define a line as identified if the position of the observed
line agrees within $0.1$\,\AA\ to the theoretical predicted
position.
 If the offset is larger than $0.1$\,\AA\, we identify a line
 feature if the characteristic shape is similar in the observations
 and computations (e.g. the line at $999.55$\,nm or $999.8$\,nm in Fig.~\ref{ident}).
In Fig.~\ref{ident}, and also in the full \feh\ atlas, we labeled
all identified lines with their quantum numbers, i.e. vibrational
assignment ($ v_u,v_l$), branch (P, Q, R), lower state rotational quantum number $J$, $\Omega$, and
in the case of blends, with their contribution to the blend. We did this for the observed
wavelength region and the complete plot is available in the online material of
this paper.
In Fig.~\ref{deltalam} we show a histogram of the residuals between the computed
line positions and the observed ones. There is no obvious systematic behavior of the scatter
with wavelength (see inset in Fig.~\ref{deltalam}).
The scatter follows a normal distribution centered at
$\Delta \lambda=0.02$\,\AA\ corresponding to $0.67$\,km s$^{-1}$. This
mean value is beyond
our spectral resolution and in general all residuals smaller than
$0.75$\,km s$^{-1} (\sim 0.025$\,\AA\ ) are not significant since they are
also smaller than the accuracy of the wavelength calibration.
From the investigation of the line positions, we saw that the
fraction of lines for which the residuals are
smaller than a certain range are distributed as shown in the right
inlay of Fig.~\ref{deltalam}.
This plots shows that $\sim 80$\,\% of the line features could be
identified by their positions which do not deviate by more than $0.1$\AA\
from the predictions.
In the cases where the residuals are
larger than $0.025$\,\AA\, the uncertainties in the molecular
constants which were used to compute the \feh\ line list are responsible for
these deviations.

For our analysis we use only lines with $W_{\lambda} > 2$\,m\AA,
which approximately describes a line with $2$\,\% relative flux and a
FWHM of $0.1$\,\AA. We ignore lines with smaller contributions because
their intensities are similar to the noise level. However, only
$167$ lines out of $1359$ have equivalent widths less than $2$\,m\AA.

\subsubsection{Vibrational bands}
\begin{figure}[!h]
\includegraphics[width=0.45\textwidth,bb = 30 30 540 380]{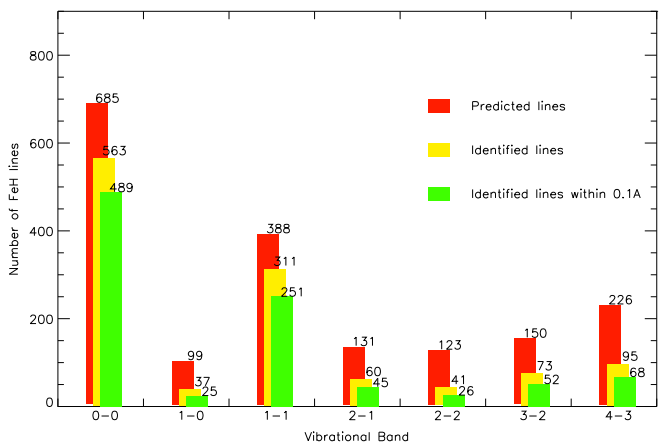}
\caption{Histogram of predicted and identified \feh\ lines in the GJ1002 spectrum.}
\label{linestat}
\end{figure}

From Fig.~\ref{FeH_all} we expect that the dominant vibrational bands are $(0,0)$ and $(1,1)$
from the $\Delta v=0$ sequence, and $(3,2)$ and $(4,3)$ from the $\Delta v=1$ sequence.
In Fig.~\ref{linestat} we present a histogram with the number of
identified lines for each vibrational band and shows that the expected
vibrational bands are present. We also show the number of
possible lines from theory with $W_{\lambda} > 2$\,m\AA\ in the observed wavelength region. This
number is based on line by line computation of $W_{\lambda}$.
In this histogram the bars in the foreground take into account only the lines
 identified by their positions.
\subsubsection{Coincidence-by-Chance Method}
For the lines with residuals smaller than $0.1$\,\AA\, we compute the
coincidence-by-chance factor $C$ from equation~\ref{moc}. In
Table~\ref{tab1} the values of $C$, $w$, $N$, and $M$ are given for
all identified bands. For the $(0,0)$ and $(1,1)$ transition, the number of
identified lines $N$ exceeds the number of coincidences by chance $C$, which is a
clear indication that these bands are present in the investigated region.
For the other
bands the situation is not so clear and we will confirm them with
further investigations.

\begin{table}
\centering
\caption{Results from the coincidence method for lines with
 $W_{\lambda} > 2$\,m\AA\ .}
\begin{tabular}{ccccc}
\hline\hline
Band & Wavelength [\AA\ ] & $C$ & $w$ [\#/\AA\ ] & $N$ ($M$) \\
\hline
$(0,0)$ & $9900-10762$  &360     & 3.73 &489~~(658)                   \\
$(1,0)$ & $9901-10759$  &52     &  3.74 &25~~(99)                    \\
$(1,1)$ & $9941-10759$  &208     & 3.71 &251~~(388)              \\
$(2,1)$ & $9931-10706$  &68     & 3.76 &45~~(131)                     \\
$(2,2)$ & $10480-10764$ & 65    & 3.81 &26~~(123)                \\
$(3,2)$ & $9910-10764$  &79     & 3.78 &52~~(150)                    \\
$(4,3)$ & $9905-10673$  &119     & 3.75 &68~~(226)                \\
\hline
\end{tabular}

\label{tab1}
\end{table}
\subsubsection{Cross-Correlation Method}
\begin{figure}[!h]
\includegraphics[width=0.5\textwidth,bb = 15 30 580 795]{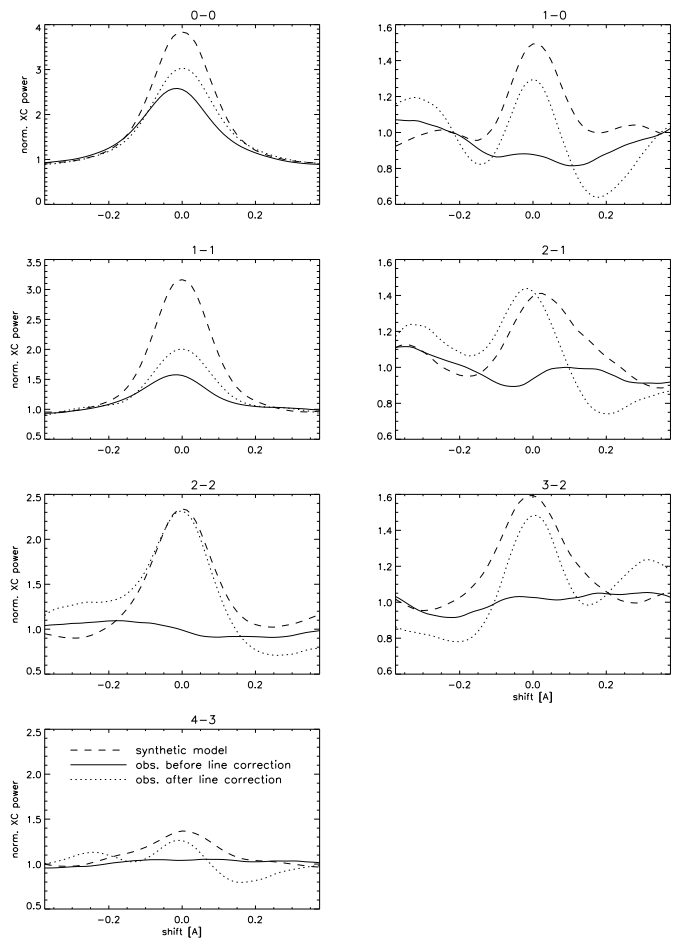}
\caption{Cross-correlation curves for different vibrational bands.}
\label{residintense}
\end{figure}

We use cross-correlation techniques to investigate the agreement
between the theoretical line list and the observed spectra.
As a reference and a test, we cross-correlate a computed
spectrum from this theoretical line list, which is broadened by an instrumental
profile with a resolving power of $70\,000$, with the original line list
\citep[e.g.][]{1995MsT..........1F,1998A&AS..129..435F}.
To be specific, we vary the
theoretical positions with steps of $\Delta \sigma=0.0125$\,\AA\ in a
range of $0.375$\,\AA\ and measure the relative intensity, at the
different positions, weighted with
$W_{\lambda}$ of the line and integrate over all lines. We then normalize the results
with the number of lines in the line list. We do this for all lines in the vibrational
bands which are found in the M dwarf spectra (see
 Fig.~\ref{residintense}). If a vibrational band is present, a peak around
zero appears above the noise produced by random
coincidences with other lines.

We produce three different curves. For the first one, we use all
possible lines from the original line list for comparison with the observed stellar spectrum
(solid line in Fig.~\ref{residintense}).
For the second curve we compute a reference curve by cross correlating the original line list with
 a synthetic spectrum computed from it (dashed line in
Fig.~\ref{residintense}). This case produces the maximum possible correlation.
A third curve is produced using the corrected line list containing only identified \feh\
lines and we compare it with the observed stellar spectrum (dotted line
in  Fig.~\ref{residintense}).

As expected, for the synthetic spectrum with the
theoretical line list (dashed line), all bands show clear peaks above the noise.
For the observed spectrum (solid line) and original line list, the $(0,0)$ and
$(1,1)$ vibrational bands show clear peaks above the noise, which is in agreement with
the coincidences by chance values in Table~\ref{tab1}. After improving
the FeH line list (dotted line), all bands show peaks above the noise
similar to the peaks obtained in the reference case from
cross-correlating the theoretical line list with the
computed spectra. The original theoretical line positions for the
$(1,0)$, $(2,1)$, $(2,2)$, $(3,2)$, and $(4,3)$ bands are not
accurate enough to show significant peaks in the cross correlation.
To confirm our identifications, we also use the line strength in the next section.

\subsubsection{Line Intensity Method}
\label{lim}
\begin{figure*}
\centering
\includegraphics[width=0.9\textwidth,bb = 15 30 580 800]{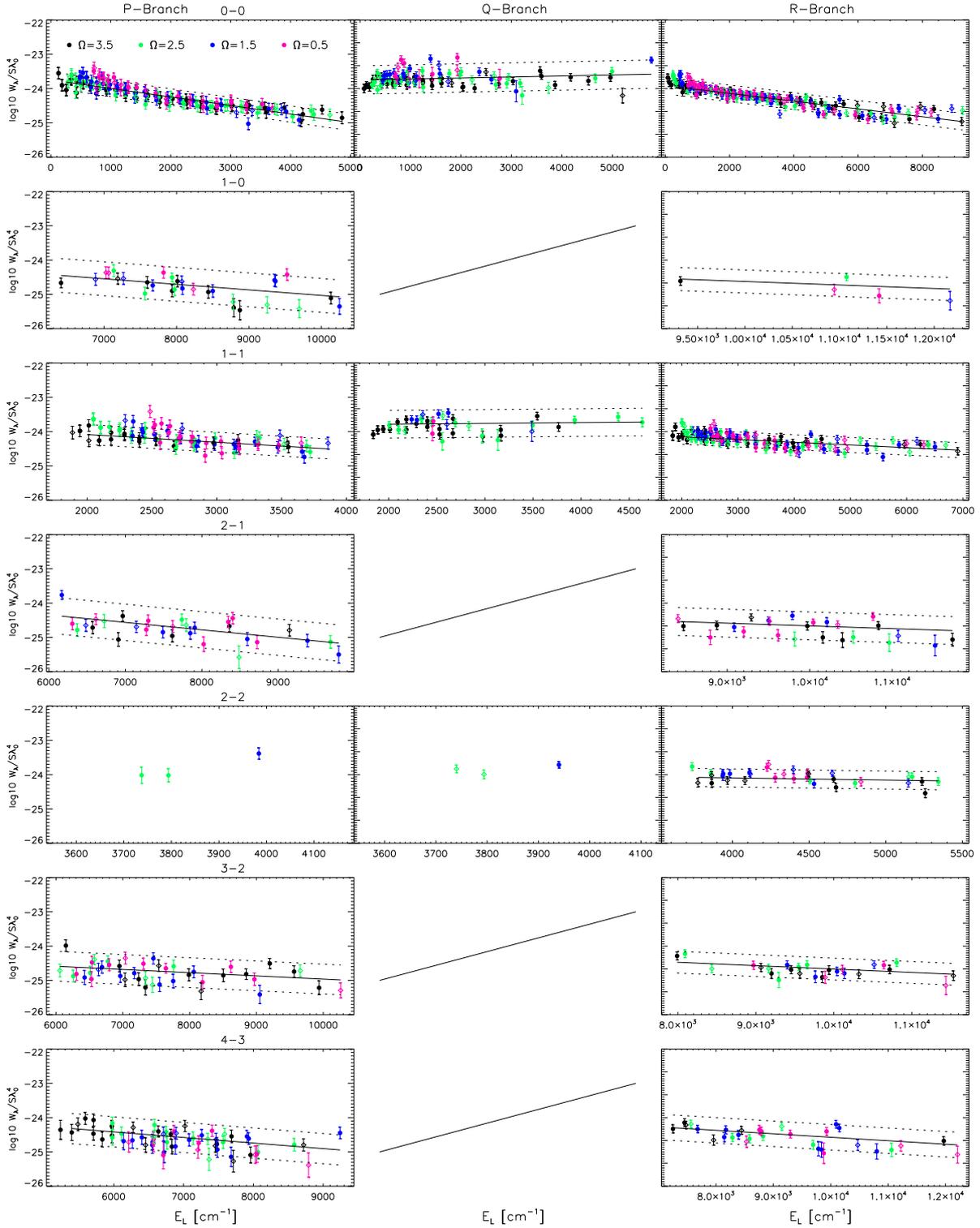}
\caption{Logarithm of observed equivalent width and theoretical line strength against
lower level energies. In each plot, a linear fit to the data is shown. Different
colors of the data points belong to different $\Omega$ values. Dots represent lines
which are identified within $0.1$\,\AA\ and diamonds lines which differ more than $0.1$\,\AA\
from the theoretical position. The dotted lines represent the three $\sigma$ scatter around the fit}
\label{ews}
\end{figure*}

We use the method described in Sec.\ref{methods} to test if the
tentatively identified observed \feh\ lines can be identified with the
theoretical ones. We plot $\log_{10}{W_{\lambda}/S\lambda_0^4}$
against $E_0$ (see equation~\ref{logS}) for the identified vibrational
bands, branches, and $\Omega$, using an appropriate estimated error
(see Fig.~\ref{ews}).

\begin{figure}
\centering
\includegraphics[width=0.5\textwidth,bb =  14 28 580 410]{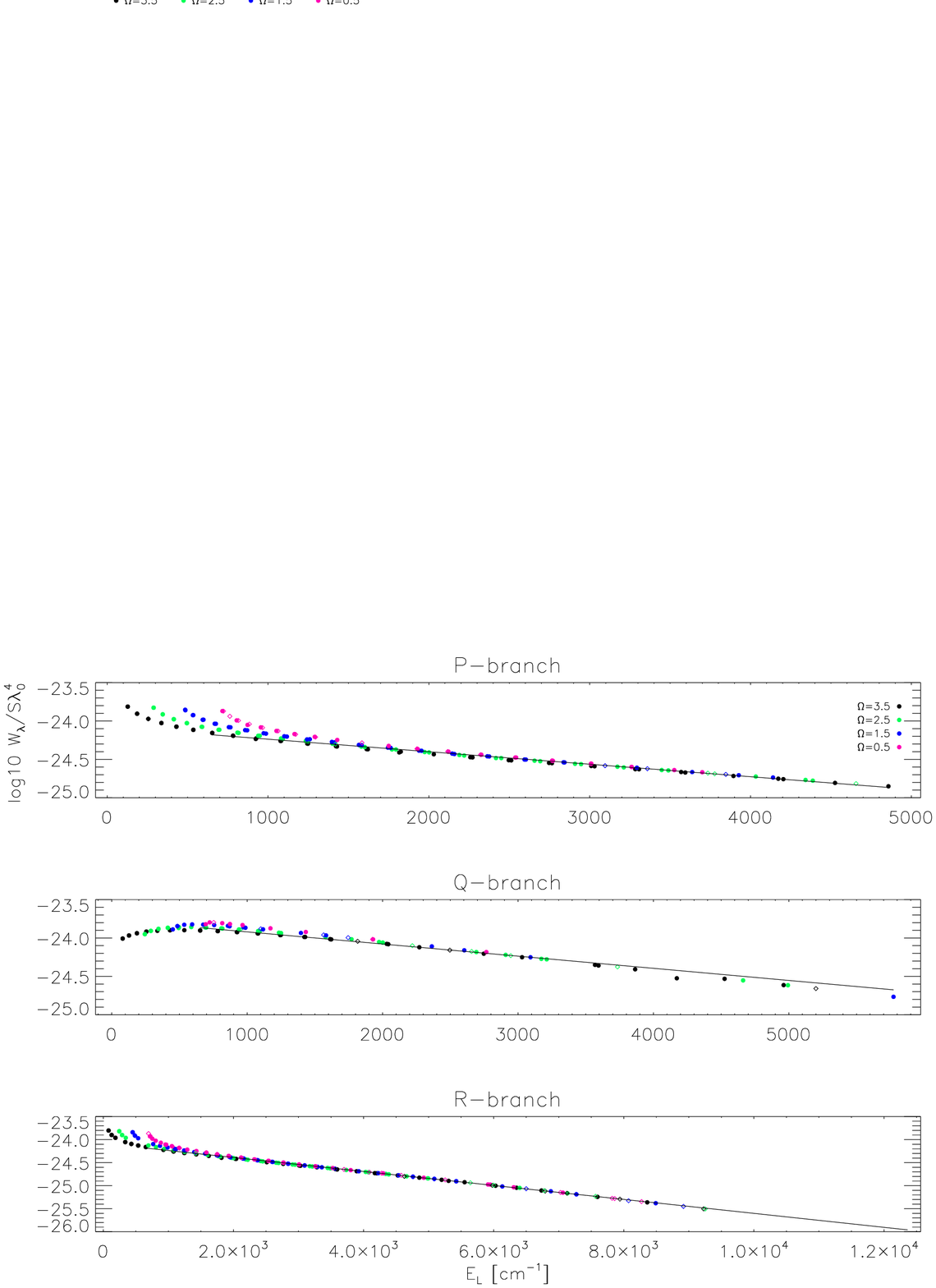}
\caption{Logarithm of computed equivalent width and theoretical line strength against
lower level energies for the $(0,0)$ band. In each plot a linear fit to the data is shown. Different
colors of the data points belong to different $\Omega$ values. Dots represent lines
which are identified within $0.1$\,\AA\ and diamonds lines which differ more than $0.1$\,\AA\
from the theoretical position.}
\label{compews}
\end{figure}

\begin{figure}
\centering
\includegraphics[width=0.5\textwidth,bb =  14 28 580 410]{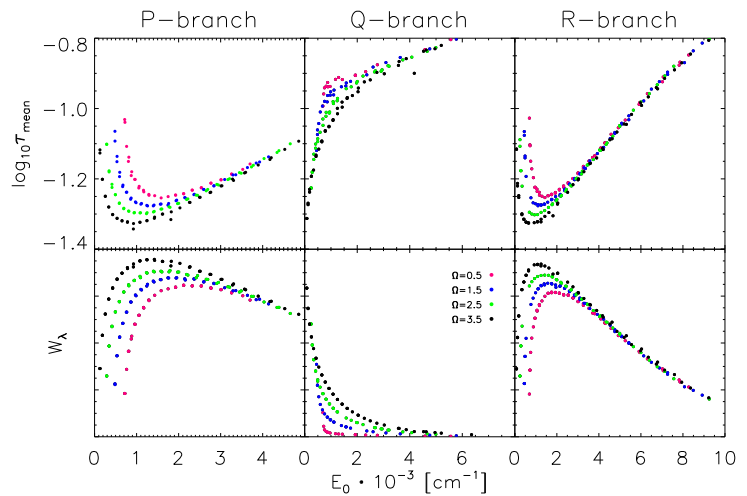}
\caption{$\log_{10}{\tau_{mean}}$ and $W_{\lambda}$ are plotted as a
 function of lower state energy for computed
 \feh\ lines from the $(0,0)$ band for the three branches. }
\label{explews}
\end{figure}

In almost all cases a linear correlation is visible. In some cases
where lines with small $J$ are present, we see a deviation from the straight line.
This deviation is expected due to different heights of formation for
the \feh\ lines. We calculated model plots with synthetic lines, which
reproduces this behavior in detail (Fig.~\ref{compews}).
The computations are very similar to the observations, which supports
the assumption of different heights of formation. To test this
assumption, we computed the contribution function for the \feh\ lines
\citep{1986A&A...163..135M} and use them to determine a weighted mean of the
continuum optical depth, $\tau_{mean}$. The latter describes a representative
atmospheric line formation depth.
In Fig.~\ref{explews}, we plotted $W_{\lambda}$ and $\tau_{mean}$
against $E_0$ for each branch from
the $(0,0)$ band. These plots show clearly that for the P- and
R-lines with low
and very high $E_0$ originate in deeper layers than the lines with
medium $E_0$. We also see that the lines with low and high $E_0$ are the ones with small
$W_{\lambda}$ because the H\"onl-London factors for P- and R-lines
are proportional to $ J_l$ and hence the lines become stronger for
large $ J_l$, but they also decrease for large $ J_l$ due to their
increasing $E_0$s. Hence, there is a maximum in $W_{\lambda}$ for medium
$ J_l$. In the case of Q-branch lines, the situation is
different. The H\"onl-London factors decrease with increasing  $ J_l$ and
hence the line strength monotonically decreases with  $ J_l$.
In this case, the lines with low $E_0$ are the
lines with the largest $W_{\lambda}$ and hence these lines are formed
in higher layers and the Q-branch shows the opposite behavior to the
P- and R-branches in the $\log_{10}{\frac{W_{\lambda}}{S\lambda^4_0}}$
plot.

\begin{figure}
\centering
\includegraphics[width=0.5\textwidth,bb =  14 28 580 410]{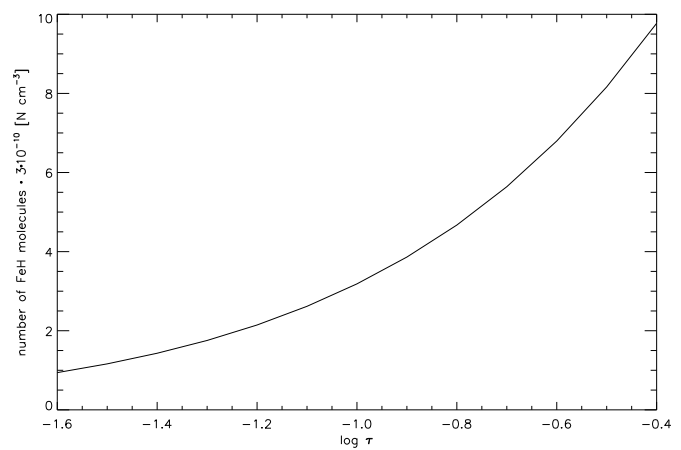}
\caption{The number of \feh\ molecules as a function of optical depth in
 the region where the \feh\ absorption lines are formed. }
\label{nfeh}
\end{figure}

We can conclude that in deeper layers, where weak lines are formed,
the equivalent width is enhanced  by
the larger number of \feh\ molecules that contribute to the absorption
(see Fig.~\ref{nfeh}). The molecule number increases towards deeper layers
due to the higher overall density
even though the concentration of \feh\ molecules relative to
\element{Fe} and \element{H} decreases. Towards
higher temperatures in deeper layers or in hotter stars, the number
of \feh\ molecules would decrease again due to the ionisation of
\element{Fe} which is then no longer available for the formation of the
molecule. In our case, this behaviour results
in a deviation from a straight line in  Fig.~\ref{ews}


With this knowledge, we can use the line strength to confirm the identification of
\feh\ .
 In Fig.~\ref{ews}, clearest correlations are found for the
$(0,0)$ and $(1,1)$ bands which have also the
largest number of lines, but also for the P-branch of the $(1,0)$ and
$(2,1)$ transitions and the R-branch of the $(2,2)$ transition. The $(3,2)$
and $(4,3)$ transitions show also linear correlations but with a larger
scatter. This result is a strong indication that all the identified bands
are present in the observed spectra.
Although the Q-branches of the $(0,0)$ and $(1,1)$ bands show no clear
linear dependence for large $E_0$, we expect them to be present and correctly
identified, since they show the expected downward trend for
small energies. The scatter towards higher energies is much larger than in the
other branches and there are only a few measurements. The
slope seems to be positive, but we could expect it to be negative if
we would have more data points.

We fit the data points linearly for lines with $J_l > 7$ (to avoid the
region where the influence of variable \feh\ number is too strong)
and use the difference of the data
points from the fit as a measure for the confidence of an identified
line.
A line is classified as ``P'' if the difference between
$\log_{10}{\frac{W_{\lambda}}{S\lambda^4_0}}$ and the linear fit is smaller than
three times the standard deviation $\sigma$ of the scatter around the fit.
But we assume that lines with $J \le 7$ are also present, since they show
exactly the expected behavior.
It is classified as ``Pb'' if the deviation is greater than $\sigma$ and the
data point is above the linear fit and as ``R'' if it is below.
It is classified as ``Q'' if the line could not be investigated
because we only investigated lines with $\Delta \Omega=0$.
We give a list of all identified lines with quantum numbers and
corrected wavelength in Table~\ref{FeH_table}. This
list is explained in more detail in the appendix.
\begin{table*}
\centering
\caption{\feh\ molecular data of the identified lines. The columns
 are described in more detail in the appendix.}
\begin{tabular}{cccccccccccccccc}
\hline\hline
$\lambda_{obs}$ & $\lambda_{theo}$ & $v_l$ & $v_u$ &$ \Omega_l$& $ \Omega_u$ & $J_l$ & $J_u$ & $B$  & $A$ &$s_A$ & $E_l$ & $\Delta \lambda$ & blend & class & comment  \\
\hline
9900.4846 &  9900.4902 & 0 & 0 &   3.5 &   3.5 &  17.5 &  18.5 & 3 &   470924.44 & 0.8309 & 0.225 &  0.0056 & 1.000 & P &   \\
9901.1049 &  9901.1175 & 0 & 0 &   3.5 &   3.5 &  14.5 &  15.5 & 3 &   462792.47 & 1.0695 & 0.154 &  0.0126 & 1.000 & P &   \\
9901.4704 &  9901.4116 & 0 & 1 &   3.5 &   3.5 &  32.5 &  31.5 & 1 &   133973.65 & 0.3378 & 0.794 & -0.0588 & 1.000 & P &   \\
9903.9631 &  9903.9809 & 0 & 0 &   3.5 &   3.5 &  18.5 &  19.5 & 3 &   472886.88 & 1.1484 & 0.252 &  0.0178 & 0.998 & P &   \\
9904.9843 &  9904.9913 & 0 & 0 &   3.5 &   3.5 &  13.5 &  14.5 & 3 &   458947.25 & 1.4267 & 0.134 &  0.0070 & 1.000 & P &   \\
9905.8556 &  9905.8940 & 0 & 0 &   3.5 &   3.5 &  15.5 &  16.5 & 3 &   466010.13 & 1.1121 & 0.177 &  0.0384 & 0.969 & P &   \\
9905.8556 &  9905.8940 & 3 & 4 &   3.5 &   3.5 &   5.5 &   4.5 & 1 &    93477.74 & 1.1121 & 0.649 &  0.0384 & 0.026 & P &   \\
$\vdots$  &  $\vdots$  & $\vdots$  & $\vdots$  & $\vdots$  & $\vdots$  & $\vdots$  & $\vdots$  & $\vdots$  & $\vdots$  & $\vdots$ & $\vdots$ & $\vdots$ & $\vdots$  &  $\vdots$  & $\vdots$  \\
\hline
\end{tabular}
\label{FeH_table}
\end{table*}

\subsection{Corrections to the line strengths}
If we want to use equation~\ref{Ascale} to correct for the differences in line depth, we have
to match the stellar parameters as closely as possible.
These parameters are basically effective temperature, surface gravity
and chemical composition,
as well as van der Waals broadening constants,  whose influence
 become significant
at these low temperatures. Van der Waals broadening is computed
with Uns\"old's hydrogenic approximation, and an enhancement factor is used model
the line wings correctly. Since for \feh\ no enhancement factor is reported, we need to
determine one.
For M-type stars, an assumed surface gravity of $\log{g}=5.0$ is
standard and the chemical composition is usually assumed as solar. To
match the strong Ti lines in the
$10\,300$--$10\,700$\,\AA\ region as well as the \feh\ lines, we
have to increase the iron abundance from $ 7.41$  to the
 \citet{1989AIPC..183....1G} value of $ 7.63$.
We use this scaling as
a parameter and do not claim this to be the actual iron abundance of GJ1002.
The free parameters for GJ1002 are now $\teff$, van der Waals
broadening enhancement factor (which we call from now on $\beta_{vdW}$),
and the instrumental resolving power, which we use as a fitting parameter to account for possible additional
rotational broadening. These three parameters are strongly correlated.
We create $\chi^2$ maps to determine the most likely combination
that matches the observed spectra best.

\subsubsection{$\chi^2$ maps}

\begin{figure}[!h]
\includegraphics[width=0.5\textwidth,bb = 15 30 580 800]{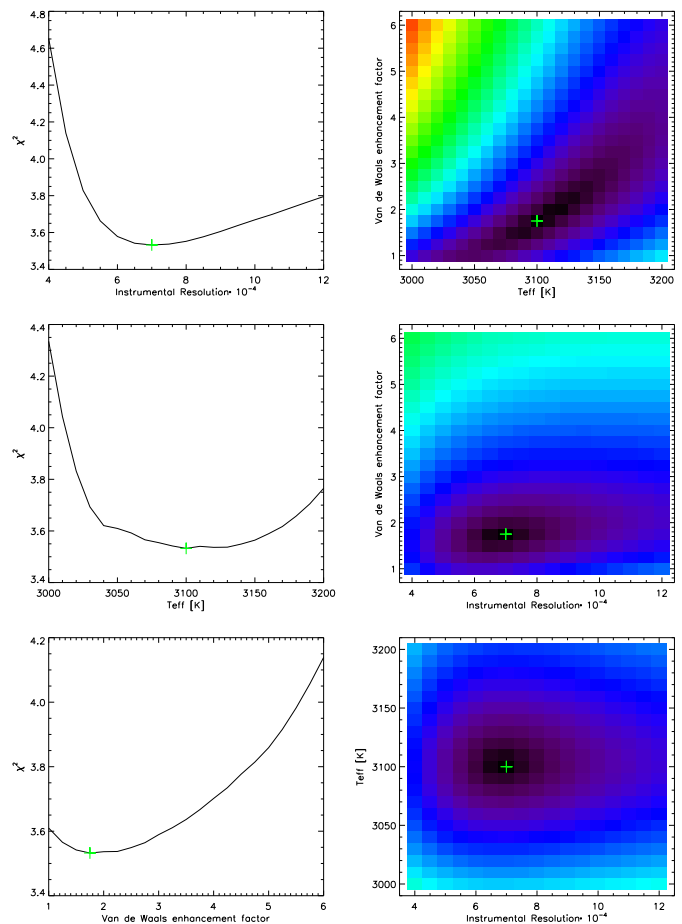}
\caption{Left: Minima in $\chi^2$ plots for the resolving power, $\teff$, and
$\beta_{vdW}$ (from top to bottom).
Right:$\chi^2$ maps for the three parameter combinations,
$\beta_{vdW}$ $-$ $\teff$, $\beta_{vdW}$ $-$
resolving power, and resolving power $-$ $\teff$
(from top to bottom). The $\chi^2$ maps are computed for the minimum
value of the leftover parameter on the left hand side.}
\label{vdwchimap}
\end{figure}

For the comparison between observation and computation we chose a region where the original
line list fits best, the lines are strongest and hence the influence of van der Waals
broadening is largest. We selected the first $100$\,\AA\ from the
$(0,0)$ band
head at $9900$\,\AA\ .
 For the computations we use our new line list with corrected positions
and include also the identified atomic lines.
To create the $\chi^2$ maps for the three combinations of parameters,
we search for the minimum for each parameter (light cross in left plots in Fig.~\ref{vdwchimap})
and use this value to construct the
$\chi^2$ map for the other two parameters.
The $\chi^2$ maps (right hand side in Fig.~\ref{vdwchimap}) yield a consistent picture
of the parameter combinations for the spectra of GJ1002.

The most likely parameters for effective temperature, resolving power,
and $\beta_{vdW}$ are $3100$\,K, $70\,000$
and $1.75$, respectively. It is obvious, however, that the $\chi^2$ curves
show regions with broad minima which would allow for variations in the
derived parameters.

Since the observed spectra are supposed to have a resolving power of
$100\,000$, the difference from our determined resolving power stems probably from rotation,
which is measured to be lower then $3$\,km s$^{-1}$, but not necessarily zero.
The difference in resolving power results in $v\sin{i} \approx 1.3$\,km s$^{-1}$ at a
wavelength of $10\,000$\,\AA\ .

An independent constraint for the instrumental resolving power and effective temperature
is given by the Ti lines, which are strong and distributed
over a wide wavelength range in the spectra. The computation of these lines with the
parameters and broadening constants given by VALD fits the
observations within $5$\,\% for the line depth. This gives us
confidence that the decreasing line depth of the \feh\ lines with
increasing wavelength is a real feature and not due to normalisation effects.

\subsubsection{Einstein A values}
\begin{figure}[!h]
\includegraphics[width=0.5\textwidth,bb = 30 30 600 410]{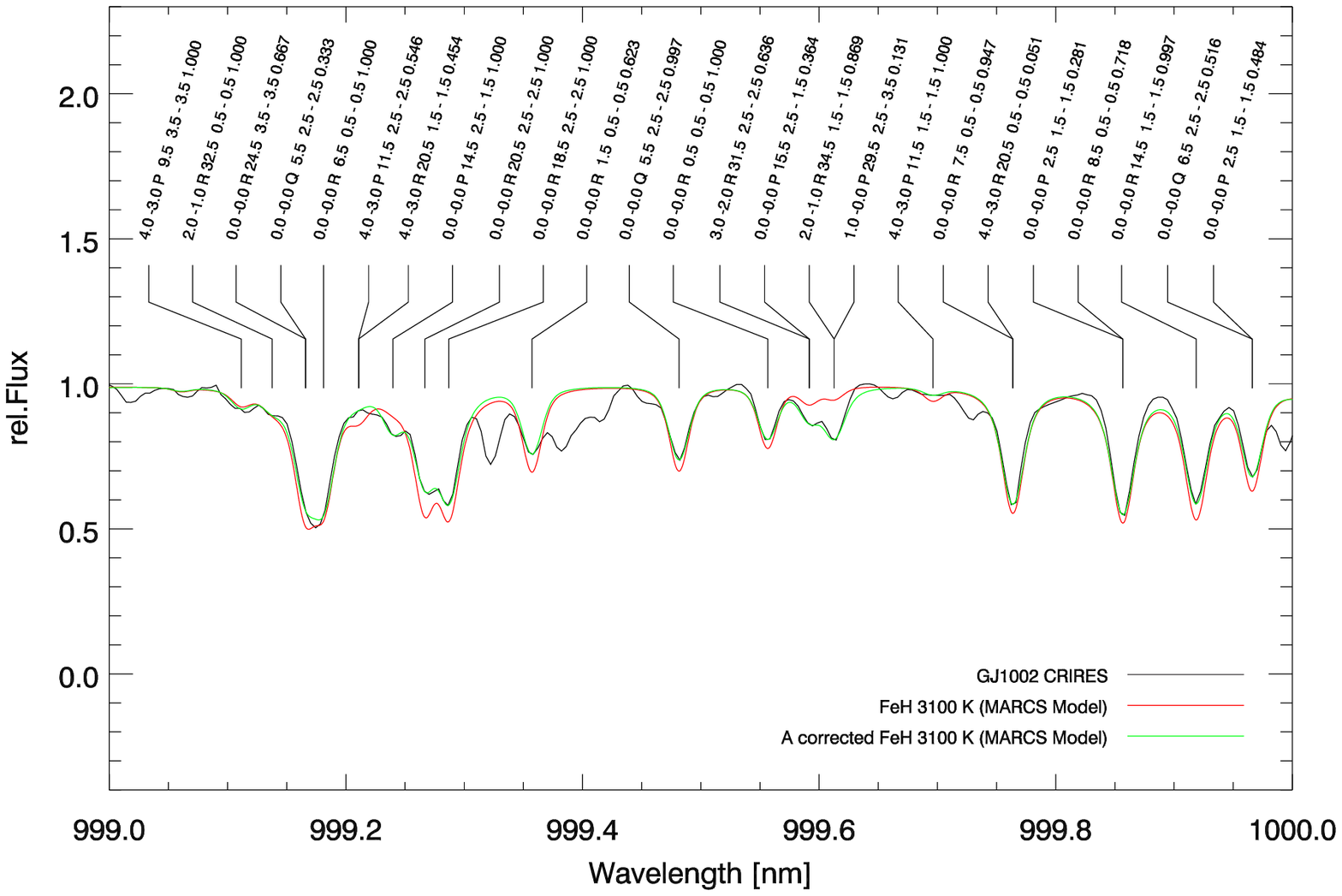}
\caption{Observed spectrum of GJ1002 (black) and computed one before A correction (red) and
after A correction (green) labeled with quantum numbers (both with
corrected positions).}
\label{Ascaling}
\end{figure}

In order to correct for the Einstein A values we use computed spectra which have already corrected
line positions. We iteratively adjust the Einstein A values because
for saturated lines the first scaling is not
sufficient. The scaling factors for each line are listed in the
Table~\ref{FeH_table} and an example for the
corrected computed spectra is shown in Fig.~\ref{Ascaling}. In order
to estimate an error, we assume an accuracy of $1\%$ for the observed
line depth, which results in an error of $\sim 3\%$ for the scaling
factors. If we furthermore assume that the line depth is modified by
an unknown blended feature by, e.g., $5\%$ an error of $\sim 16\%$
follows. The accuracy of the Einstein A scaling is also
influenced by the van der Waals enhancement factor $\beta_{vdW}$. A
change of $\pm1$ gives a mean difference of $\sim 5\%$ in the scaling
constants, but can be up to $30\%$ for individual lines, due to the
logarithmic ratio of the intensities (see
e.g. equation~\ref{Ascale}).

We obtain a good fit to the data with
the scaled Einstein A values even for some lines calculated to be very
weak. However this results in some cases in unrealistic
$\log{gf}$ values, and we assume that these weak lines are blended with
unknown components.
The scaling of line blends is a difficult problem because
it results in equal scaling factors for lines with completely
different quantum numbers. To avoid this problem one could
determine scaling factors for each branch, but we chose the simpler scheme of scaling each line.
\begin{figure}[!h]
\includegraphics[width=0.5\textwidth,bb = 30 30 600 410]{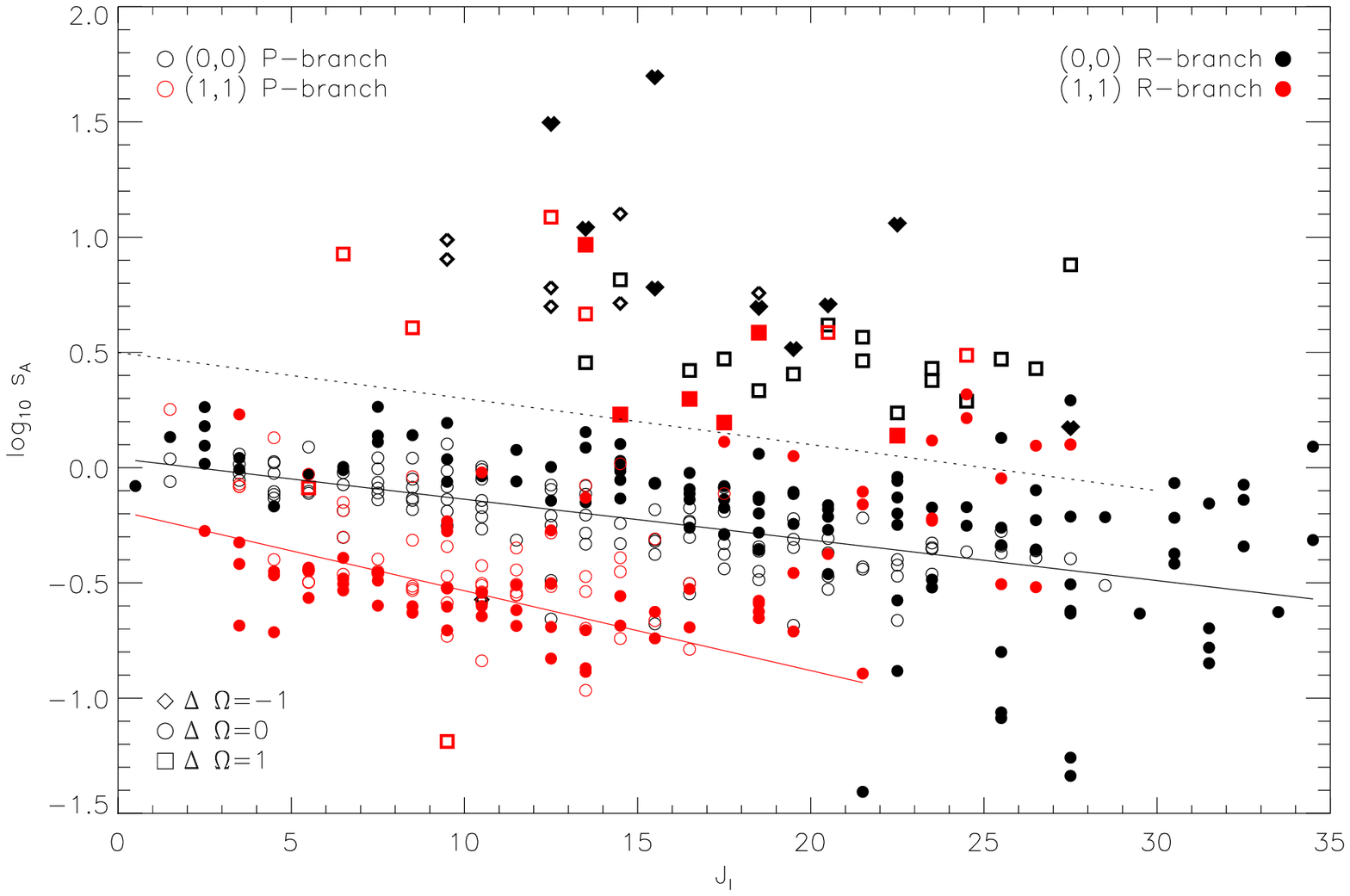}
\caption{Scaling factor for the Einstein A values against rotational
quantum number for the $(0,0)$ and $(1,1)$ bands. The plot is
truncated at $ J_l=35$ for better visibility.} 
\label{Acorrfig}
\end{figure}
We use equation~\ref{Ascale} and plot $s$ for the Einstein A values
against  $ J_l$ to search for a possible rotational dependence (Fig.~\ref{Acorrfig}).
We plot only scaling
factors for the $(0,0)$ and $(1,1)$ bands, since these are the bands
with the largest number of lines and we use only lines which
contribute more than $99$\,\% to a blended line feature in order to avoid
contributions from incorrectly scaled lines.

We find two groups of
scaling factors. One group describes a strong enhancement of the line
depths (`positive' scaling factors) and the other only a small
enhancement for low  $ J_l$ and a reduction towards lines with large
$ J_l$ (`negative' scaling factors).  We divide these two groups by a
dashed line in Fig.~\ref{Acorrfig}.
For the $\Delta \Omega=0$ transitions of the $(0,0)$ and $(1,1)$ bands
we include linear fits to the data in Fig.~\ref{Acorrfig} to indicate the slope.

The group of positive scaling factors is strongly
dominated by $\Delta \Omega \pm 1$ transitions (diamonds and
rectangles), while the group of negative scaling factors consists of only $\Delta \Omega=0$
transitions (circles). The latter scaling factors for $(0,0)$ and
$(1,1)$ bands (black and red circles, open for P-branches and filled for
R-branches) show an almost linear behaviour with  $ J_l$ and become stronger
towards larger  $ J_l$. The positive scaling factors show also two linear groups, which originate
from the $(1,1)$ band (red rectangles) and from the $(0,0)$ band
(black diamonds and rectangles). These scaling factors describe a
strong enhancement of the lines for small  $ J_l$ and become smaller
towards large $ J_l$.
All groups of scaling factors have a similar negative slope, and are only
shifted by a constant factor to higher or lower scaling factors.

 The distribution of other bands and lines
which contribute less than $99$\,\% to a blended feature gives only a
larger scatter to the data points, but does not change the basic trend
of the scaling factors.
We conclude that the  $ J_l$ dependence in the
scaling factors likely indicates shortcomings in the calculated
H\"onl-London factors. In particular, satellite branches with $\Delta
\Omega \pm 1$ are much stronger than expected for Hund's case (a).
In other words, the two $^4\Delta$ electronic states
are heavily mixed with other electronic states and the
simple Hund's case (a) behaviour anticipated for isolated electronic
states with relatively large spin-orbit splittings is not found.

\subsection{Rotational temperatures}
A rotational temperature $T_{rot}$ can be
obtained from the slope $m$ of the linear fit in the
$[\log_{10}{W_{\lambda}}/S,E_0]$
diagram described in the above section. $T_{rot}$ can be calculated from
\bq
T_{rot}=\frac{hc}{mk}\log_{10}{e}.
\label{Trot}
\eq
For the fit to the data, we use only lines with $ J_l>7$ in order to avoid
the significant influence from varying absorber number,
and neglect lines with line depth greater than $0.5$ to avoid
saturation effects.

\begin{figure}
\includegraphics[width=0.5\textwidth,bb = 20 30 580 410]{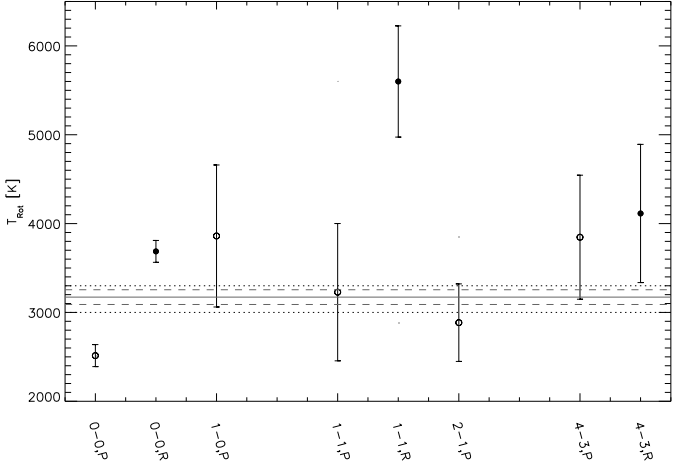}
\caption{Rotational temperatures derived from the slope of the linear fits in Fig.~\ref{ews}.
The error bars indicate the one sigma level. The dotted
lines give the expected upper and lower effective temperatures for a 5.5 M-dwarf. The grey solid
line is the weighted mean of the rotational temperatures with its one
sigma error (dashed-line).}
\label{T_rot}
\end{figure}

 Due to the large errors (derived from the uncertainty of the
slope of the linear fit in Fig.~\ref{ews}), we consider only rotational
temperatures with moderately small one sigma errors (see Fig.~\ref{T_rot}).


 We find systematically lower temperatures for the
P-branches in comparison to the R-branches which is consistent with
different heights of formation for most of the lines in a branch as seen in Fig.~\ref{explews}.

If we compute the weighted mean of the rotational
temperatures (grey solid line in Fig.~\ref{T_rot}) with the $1\sigma$-error
(grey dashed-line in Fig.~\ref{T_rot}), we
obtain $ \overline{T}_{rot}\approx3200\pm100$\,K . This is in the middle of
the expected temperature range for this spectral type ($3000$\,K--$3300$\,K, dashed-dotted
lines in Fig.~\ref{T_rot}) and is close to our estimated value of
$\sim 3100$\,K. The
main contribution to $\overline{T}_{rot}$ stems from
the P- and R-branches of the $(0,0)$ transition due to their small
uncertainties. The weighted mean of the P- and R-branches are
$ \overline{T}_{rot}^P\approx2600\pm 150$\,K and
$ \overline{T}_{rot}^R\approx3750\pm 150$\,K, respectively.

We want to point out that even for a single branch in a band, the
obtained temperature is an average over the individual excitation
temperatures for each line. The resulting temperature
depends crucially on the selection of the lines that are used. If one
uses lines with similar equivalent width and lower level energy,
then one could obtain temperatures for certain regions in the atmosphere. For this
method, however, a large number of lines is required, because otherwise the
uncertainties become too large.

We conclude that in order to use the method of rotational temperatures,
a large number of well measured lines are required to
minimize the error in the slope. Finally, the rotational temperature
can only be expected to match the effective temperature if the lines
form in a region around optical depth unity.

\section{Summary and conclusion}

We investigated the z-band region of the M5.5 dwarf GJ1002 with
high-resolution CRIRES spectra. This is the region where the $(0,0)$
vibrational band of \feh\ is present (`Wing-Ford band'). We
were able to identify the $(0,0)$, $(1,0)$,
$(1,1)$, $(2,1)$, $(2,2)$, $(3,2)$, and $(4,3)$ bands. For
confirmation of the band assignment, we used the method of coincidence,
cross-correlation techniques, and the line intensities.

For the identified lines, we applied empirical corrections to the theoretical
line positions. The deviations between observed and computed positions are
Gaussian distributed around zero. In the case of
small deviations ($<0.025$\,\AA\ ), this could be due to uncertainties
in the wavelength calibration of the observed spectra and in other
cases due to uncertainties in the molecular constants which were used
to generate the line list.  Note that these empirical wavelength adjustments 
reproduce the observed stellar spectrum, but the quantum number assignments are by no means assured.   

The method of coincidence confirms the presence of the $(0,0)$ and
$(1,1)$ bands, but for the other bands we needed cross-correlation
techniques to show their presence. Again the $(0,0)$ and $(1,1)$ bands
show clear peaks in the cross-correlation function even with an
uncorrected line list, but for the other bands it was necessary to use
corrections to the line list in order to confirm them.
We also used line intensity information by plotting
the ratio $W_{\lambda}/S\lambda_0^4$ against $E_0$ which shows
a linear behaviour if the band is present in the spectra. With this
method we could confirm the
presence of all the other bands although not uniquely for the Q-branches.

From the slope of the line in the $[W_{\lambda}/S\lambda_0^4,E_0]$ plots, it
was possible to derive excitation temperatures for rotational
transitions, which could be identified with the effective temperature of
the star if the lines are formed in the photosphere.
We showed that this method is very uncertain since the
error in the slope is high. The derived temperatures for the
individual vibrational transitions range from
$ \sim 2500$\,K to $ \sim 5500$\,K for GJ1002, but the weighted mean
$\overline{T}_{rot}\approx3200\pm100$ K is very close to the expected
temperature of an M 5.5 dwarf. However, the large error bars and
differences between P- and R-branch temperatures suggests that this
agreement may be more of a coincidence than a physical
result.

Finally we corrected the line strength in the \feh\ line list by scaling the Einstein A
values, as some of these lines show large discrepancies compared to the
observations. For this purpose it was necessary to derive the instrumental
broadening (which included the rotational broadening), effective
temperature, and an enhancement factor for the van der Waals
broadening constants. The instrumental resolving power was derived to $R=70\,000$, which
is equivalent to a rotational broadening with $vsin(i)\sim 1.3$\,km s$^{-1}$. We
also derived an effective temperature of
$\teff=3100$\,K, and a van der Waals enhancement constant of $1.75$.
The scaling factors of the Einstein A values show an almost linear
dependence on  $ J_l$, which indicates that there is likely a
problem in the calculation of the H\"onl-London factors.

With the improved identification of \feh\ lines, it is now possible to
characterize the \feh\ lines in the z-band region
(e.g. magnetically sensitive and insensitive lines, temperature
sensitivities of individual lines). Our improved line list will aid in
the identification and simulation of \feh\ lines in spectra of cool
stars.

\section*{Appendix}

\subsection*{Explanation of the FeH Table}
The Table~\ref{FeH_table} contains the observed
wavelength $\lambda_{obs}$ which are obtained with the Voigt fit and
the theoretical wavelength $\lambda_{theo}$ from the list of
\citet{2003ApJ...594..651D}. The wavelengths are in vacuum. We give
also the quantum numbers of the lines and the Einstein A values with
their scaling factors $s_A$. The lower level energy $E_l$ is given in
eV. The difference in position $\Delta
\lambda=\lambda_{theo}-\lambda_{obs}$ is also printed in the table. If the line is a
blended line, then its contribution to the blend is given as the
fraction normalized to one. If the line is not blended its blend value
is one. We give then the classification of the line, as defined in
section 3.3. We add a comment if the line is blended by an atomic
feature, or if the classification of the line did not agree with the scaling
factor of the Einstein A values.

\subsection*{Explanation of the FeH Atlas}
We plotted the whole spectrum in bins of $1$\,nm\ from $990$\,nm to
$1076.6$\,nm.
Shown are the observed spectrum of GJ1002 (black), the computed
spectrum with corrected positions (red), the computed spectrum with
corrected positions and scaled Einstein A values (green). We also
labeled all lines with $W_{\lambda} \ge 2$\,m\AA\ with quantum number
for the vibrational transition, the branch,  the lower $J$, the upper and
lower $\Omega$, and in the last position, their blend fraction. The blend fraction is unity if a line is not blended. We also labeled
the position of atomic lines with the element name below the spectrum.


\begin{acknowledgements}
SW would like to acknowledge the support from the DFG Research Training
Group GrK - 1351 ``Extrasolar Planets and their host stars''.\\
AR acknowledges research funding from the DFG under an Emmy Noether
Fellowship (RE 1664/4- 1).\\
AS acknowledges financial support DFG under grant RE 1664/4- 1 and from
NSF under grant AST07-08074.\\
Some support to PFB was provided by the NASA laboratory astrophysics program.\\
We thank P. Hauschildt and D. Shulyak for useful discussions.
\end{acknowledgements}

\bibliographystyle{aa}
\bibliography{15220_arxiv}

\clearpage
\clearpage

\end{document}